\documentstyle[prl,aps,preprint,eqsecnum]{revtex}

\def\a{\alpha}
\def\b{\beta}

\def\vta{\vartheta}

\def\bg{{\bf g}}

\def\negenspace{\kern-1.1em}\def\quer{\negenspace\nearrow}

\begin{document}

\title{Poincar\'e gauge gravity: selected topics}

\author{Yuri N.~Obukhov\footnote{On leave from: Dept. of Theoret. Physics, 
Moscow State University, 117234 Moscow, Russia}}
\address{Institute for Theoretical Physics, University of Cologne,
50923 K\"oln, Germany}

\maketitle
\begin{abstract}
In the gauge theory of gravity based on the Poincar\'e group (the semidirect
product of the Lorentz group and the spacetime translations) the mass 
(energy-momentum) and the spin are treated on an equal footing as the sources
of the gravitational field. The corresponding spacetime manifold carries
the Riemann-Cartan geometric structure with the nontrivial curvature and
torsion. We describe some aspects of the classical Poincar\'e gauge theory 
of gravity. Namely, the Lagrange-Noether formalism is presented in full
generality, and the family of quadratic (in the curvature and the torsion)
models is analyzed in detail. We discuss the special case of the spinless
matter and demonstrate that Einstein's theory arises as a degenerate model
in the class of the quadratic Poincar\'e theories. Another central point
is the overview of the so-called double duality method for constructing of 
the exact solutions of the classical field equations.   
\end{abstract}
\bigskip

\noindent PACS: 04.50.+h, 04.30.-w, 04.20.Jb. 

\section{Introduction}

In this paper, we do not aim to give an exhaustive review of
the Poincar\'e gauge theory of gravity. Correspondingly, our list of 
references is far from being complete. The interested reader can find
more details and more literature in the review papers and the books 
\cite{Blag,Hehl1,Hehl2,Hehl3,Ivan1,Ivan2,Mccrea3,Mielke3,PBO,Sard,Shapiro}.

Within the framework of the gauge approach to gravity, the {\it kinematic} 
scheme of the theory is well understood at present, see, for example,
\cite{Blag,Hehl1,Hehl2,Hehl3,Ivan2,Trautman1,Trautman2,Trautman3,PBO,Sard}. 
The latter is based on the fiber bundle formalism and the connection theory 
combined with a certain spontaneous symmetry breaking mechanism, see also 
\cite{Tseytlin,Tres,Percacci}. These aspects are
described in the references given above, and we will not discuss them here. 
Instead, we will start from the final point of the kinematic scheme 
which introduces the Riemann-Cartan geometry on the spacetime manifold 
with the coframe and linear connection as the fundamental
variables which describe the gravitational field. 

However, the {\it dynamic} aspects of Poincar\'e gauge gravity have been 
rather poorly studied up to now. The choice of the basic Lagrangian of the 
theory still remains an open problem, and this, in turn, prevents a detailed 
analysis of possible physical effects. As a first step, 
one can use a correspondence principle. It is well known that Einstein's 
general relativity theory is satisfactorily supported by experimental tests 
on the macroscopic level. Thus, whereas the gravitational gauge models 
provide an alternative description of the gravitational physics in the microworld, 
it is natural to require their correspondence with the general relativity at 
large distances. In other words, of particular interest is the limit of
spinless matter and the possibility of a reduction of the field equations to the
(effective) Einstein theory. In this paper we will address these questions.

The structure of the paper is as follows. In Sec.~\ref{Lagrange} we fix the
notation and set the general framework for the Lagrangian theory of the 
interacting gravitational and material fields. Sec.~\ref{matlag} is devoted
to the analysis of the symmetries of the action of matter. The energy-momentum
and the spin currents are introduced. They satisfy the Noether identities derived
in Sec.~\ref{noeth}. The gravitational field momenta are introduced in 
Sec.~\ref{gravlag} and the corresponding Noether identities are obtained for 
the diffeomorphism and the local Lorentz invariance of the action of the Poincar\'e 
gravitational field. The general system of the gravitational field equations
is established in Sec.~\ref{fieldeqs}. Then, in Sec.~\ref{spinless} we discuss
the limit of spinless matter. After these preparations, we turn attention to the
general quadratic Poincar\'e gauge models and derive the corresponding field
equations in Sec.~\ref{quadratic}. This class of models is important
because of its similarity with the Yang-Mills theories of the internal 
symmetries. In Sec.~\ref{Einsteinlimit} we demonstrate that
the Einstein theory can be interpreted as a special degenerate case of the
quadratic Poincar\'e gauge theory. The notion of double duality for the 
tensor-valued 2-forms is introduced in Sec.~\ref{doubleduality} and the 
double duality properties of the irreducible parts of the Riemann-Cartan 
curvature are derived. These properties underlie the {\it double duality ansatz}
(DDA) which proved to be a very powerful tool for solving the classical 
gravitational field equations in vacuum and with the nontrivial matter sources. 
The latter, under the DDA assumption, reduce to the effective Einstein theory. 
As an example, in Sec.~\ref{kinky} we describe the exact {\it kinky-torsion} 
solution of the equations for the coupled gravitational and the Higgs-type
scalar fields. Finally, in Sec.~\ref{vacuum} we demonstrate that the Einstein 
spaces are the generic {\it torsion-free} solutions of the quadratic Poincar\'e 
gauge models. We collect the technical details of the computations in the 
three appendices. 
 
Our basic notation and conventions are those of the ref. \cite{Hehl2}. In particular,
the Greek indices $\alpha,\beta,\dots = 0,\dots,3$, denote the anholonomic 
components (for example, of a coframe $\vartheta^\alpha$), while the Latin indices 
$i,j,\dots =0,\dots, 3$, label the holonomic components ($dx^i$, e.g.). The 
volume 4-form is denoted $\eta$, and the $\eta$-basis in the space of exterior
forms is constructed with the help of the interior products as $\eta_{\alpha_1
\dots\alpha_p}:= e_{\alpha_p}\rfloor\dots e_{\alpha_1}\rfloor\eta$, $p=1,\dots,4$.
They are related to the $\theta$-basis via the Hodge dual operator $^\star$, for
example, $\eta_{\alpha\beta} = {\frac 12}\,{}^\star\left(\vartheta_\alpha\wedge
\vartheta_\beta\right)$.

\section{Lagrange formalism for the gauge gravity theory}\label{Lagrange}

The Poincar\'e gauge gravitational potentials are the local coframe 1-form
$\vartheta^\alpha$ and the 1-form $\Gamma_\alpha{}^\beta$ of the metric-compatible
connection.  Moreover, the metric structure $\bg$ is defined on the
spacetime manifold $M$. We can describe the latter explicitly by using the metric
components $g_{\alpha\beta}:=\bg(e_\alpha,e_\beta)$ calculated on the
frame $e_\alpha$ dual to $\vartheta^\alpha$. The compatibility of the metric 
and the connection is formulated in terms of the vanishing nonmetricity:
\begin{equation}
Q_{\alpha\beta}:=-Dg_{\alpha\beta} = -dg_{\alpha\beta} + 
\Gamma_{\alpha}{}^\gamma g_{\gamma\beta} + \Gamma_{\beta}{}^\gamma 
g_{\alpha\gamma}=0.\label{zeroQ}
\end{equation}
Similarly to the freedom of the choice of the local coordinates, we notice that
there is no a priori any reason to confine oneself to the case of the 
orthonormal frames for which $g_{\alpha\beta}$ is reduced to $o_{\alpha\beta}
= {\rm diag}(+1,-1,-1,-1)$. Mathematically, the vanishing of the nonmetricity 
(\ref{zeroQ}) means that the
symmetric part of the connection is completely determined by the metric,
\begin{equation}
\Gamma_{(\alpha\beta)}={\frac 1 2}dg_{\alpha\beta},\label{Gsym}
\end{equation}
and thus only the antisymmetric piece
\begin{equation}
\Gamma_{[\alpha\beta]}
\end{equation}
is the true independent (of the coframe and the metric) gravitational potential.

With the help of the local general linear transformation
\begin{equation}
e'_\alpha = \Lambda(x)_\alpha{}^\beta e_\beta, \quad\quad
\left({\rm hence}\quad \vartheta^{\prime\alpha}=\Lambda^{-1}
(x)_\beta{}^\alpha\vartheta^\beta\right),\label{genlin}
\end{equation}
one may go to the {\it gauge} in which $g_{\alpha\beta}$ is {\it constant}
(but not necessarily equal to $o_{\alpha\beta}$), thus eliminating the
symmetric part of connection, $\Gamma_{(\alpha\beta)}=0$. The well known
examples of such a gauge are provided by the null (or Newman-Penrose) and 
semi-null frames which are not orthonormal. 

It is worthwhile to note that (\ref{zeroQ}) guarantees the skew symmetry
of the curvature 2-form,
\begin{equation}
R_{\alpha\beta}=R_{[\alpha\beta]}
\end{equation}
which is true for all choices of the frame. Indeed, taking the covariant
derivative of (\ref{zeroQ}) and using the Ricci identity, we find
\begin{equation}
2R_{(\alpha\beta)}=-DDg_{\alpha\beta}=DQ_{\alpha\beta}=0.
\end{equation}

The material fields may be scalar-, tensor- or spinor-valued forms of
any rank, and we will denote them collectively as $\Psi^A$, where the 
superscript $\scriptstyle A$ indicates the appropriate index (tensor
and/or spinor) structure. 
In many cases we will suppress the generalized 
index $\scriptstyle A$. We will assume that the matter fields $\Psi^A$ 
belong to the space of some (reducible, in general) representation of the 
Lorentz subgroup $SO(1,3)$ of (\ref{genlin}) which is, as usually, defined 
by the condition of invariance of the metric,
\begin{equation}
\Lambda(x)_\alpha{}^\beta\in SO(1,3)\Longleftrightarrow 
\Lambda(x)_\alpha{}^\mu\Lambda(x)_\beta{}^\nu g_{\mu\nu} =
g_{\alpha\beta}.\label{so13}
\end{equation}
For infinitesimal transformations,
\begin{equation}
\Lambda(x)_\alpha{}^\beta = \delta_\alpha^\beta + 
\omega_\alpha{}^\beta,\label{lor0}
\end{equation}
we find the standard skew symmetry condition
\begin{equation}
\omega_{\alpha\beta} = - \omega_{\beta\alpha},\label{lor1}
\end{equation}
and hence the corresponding transformation of the material fields is 
given by
\begin{equation}
\Psi^{\prime A}= \Psi^A + \delta\Psi^A,\quad\quad
\delta\Psi^A=-\omega_\alpha{}^\beta \rho(L^\alpha{}_\beta)^A{}_B\Psi^B,
\end{equation}
where $L^{\alpha\beta}=L^{[\alpha\beta]}$ are the Lorentz group generators,
and $\rho$ describes the matrix representation of the Lorentz group. For
brevity, we will use more compact notation, $\rho^{\alpha A}_{\beta B}:=
\rho(L^\alpha{}_\beta)^A{}_B$. 

The dynamics of the theory is determined by choosing a scalar-valued
Lagrangian 4-form 
\begin{equation}\label{lagr0}
L_{\rm tot}=L(g_{\alpha\beta},\Psi^A,d\Psi^A,\vartheta^\alpha,
   \Gamma_\alpha{}^\beta) + V(g_{\alpha\beta},\vartheta^\alpha, 
   \Gamma_\alpha{}^\beta, d\vartheta^\alpha, d\Gamma_\alpha{}^\beta)\,, 
\end{equation} 
(plus possible surface term $V_{\text{surface}}$ which is often necessary
for ensuring the correct boundary conditions), where $L$ is 
the material and $V$ the gravitational Lagrangian. Note that the Lagrangian
does not contain derivatives of the metric in view of (\ref{Gsym}). The field 
equations are then found by requiring that the action integral
\begin{equation}\label{act0}
   W=\int\limits_U\, L_{\rm tot}\,, 
\end{equation}
should have a stationary value for arbitrary variations $\delta\Psi^A$,
$\delta\vartheta^\alpha$, $\delta\Gamma_\alpha{}^\beta$ of $\Psi^A$, 
$\vartheta^\alpha$ and $\Gamma_\alpha{}^\beta$, which vanish on the boundary 
$\partial U$ of an arbitrary 4-dimensional region $U\subset M$ of spacetime. 
In other words, we require
\begin{equation}
   \delta W = \int\limits_U\, \delta L_{\rm tot} = 0\,, 
\end{equation}
under the stated conditions. Noether identities will follow from
the requirement that $L$ is scalar-valued 4-form with respect to 
frame transformations (\ref{genlin}) and, since there is no explicit 
coordinate dependence, it is invariant under arbitrary diffeomorphisms. 

A separate remark is necessary about the status of the metric $g_{\alpha\beta}$ 
which can enter explicitly in (\ref{lagr0}). At the first sight it may 
seem to be an additional dynamical variable. However, it is not. The variation 
with respect to the metric vanishes as a result of the Noether identities (see 
below). As a matter of fact, this result is fairly clear if one recalls that, 
using the invariance of (\ref{lagr0}) under the change of the frame and local 
coordinates, we may completely eliminate the metric by choosing the convenient 
gauge, e.g., $g_{\alpha\beta}= o_{\alpha\beta}$.

Having thus outlined the programme that we intend to pursue, it remains
to carry out this programme step by step. We do this initially
without specifying the precise form of $L$ or $V$. General relativity, the 
Einstein-Cartan theory and the quadratic gauge theory enter as special cases
of $V$. As a historic remark, let us mention that the general aspects of the
Lagrangian approach for the Poincar\'e gauge gravity were analyzed, for 
example, in \cite{Frolov,Katanaev,Szczyrba} and in the review papers
\cite{Hehl1,Hehl2}.  

\section{Material sources}\label{matlag}

We assume that the material Lagrangian $4$--form $L$ depends most generally 
on $\Psi$, $d\Psi$, the gravitational potentials $\vartheta^{\alpha}$,
$\Gamma_{\alpha}{}^{\beta}$, and the metric $g_{\alpha\beta}$ which may have
nontrivial components in an arbitrary frame. According to the {\it minimal 
coupling} prescription, derivatives of the gravitational potentials are not 
permitted. We usually adhere to this principle. However, the {\it Pauli type terms}
and the Jordan--Brans--Dicke type terms may occur in the phenomenological models 
or in the context of a symmetry breaking mechanism. Also the Gordon decomposition 
of the matter currents and the discussion of the gravitational moments necessarily
requires the inclusion of the Pauli type terms, see \cite{Hehl4,Fermion}. Therefore, 
we develop our Lagrangian formalism in a sufficient generality in order to cope with 
such models by including in the Lagrangian also the derivatives 
$d\vartheta^{\alpha}$, and $d\Gamma_{\alpha}{}^{\beta}$ of the gravitational 
potentials:
\begin{equation} 
L= L( g_{\alpha\beta}\,,\vartheta^\alpha\,, d\vartheta^\alpha\,,
\Gamma_{\alpha}{}^{\beta}\,, d\Gamma_{\alpha}{}^{\beta}\,,\Psi^A, d\Psi^A)\,.
\end{equation}
As a further bonus, we can then also read off the Noether identities for the 
gravitational gauge fields by considering the subcase $\Psi^A=0$. 

One consequence of the invariance of $L$ under frame transformations
(\ref{genlin}) is that it can be recast in the form 
\begin{equation}\label{K}
L=L(g_{\alpha\beta}, \Psi^A, D\Psi^A, \vartheta^\alpha, 
T^\alpha, R_\alpha{}^\beta)\,. 
\end{equation} 
That is, the exterior derivative $d$ and the connection form 
$\Gamma_\alpha{}^\beta$ can only occur in the combination which gives the {\it 
covariant\/} exterior derivative 
\begin{equation}
D=d +\rho_{\mu B}^{\nu A}\,\Gamma_\mu{}^\nu\wedge,
\end{equation}
whereas the derivatives of the coframe and connection can only appear via
the torsion and the curvature 2-forms. In order to see this, we use the fact 
that at any given event $x$ there exists a
frame such that $\Gamma_\alpha{}^\beta=0$ at $x$. Then, in that frame, 
\begin{equation}
L(g_{\alpha\beta}, \Psi^A, d\Psi^A, \vartheta^\alpha, d\vartheta^\alpha\,,
\Gamma_{\alpha}{}^{\beta}\,, d\Gamma_{\alpha}{}^{\beta})\stackrel{*}{=}
{\hat L}(g_{\alpha\beta}, \Psi^A, D\Psi^A,\vartheta^\alpha, T^\alpha, 
R_\alpha{}^\beta)\,,\label{normL}
\end{equation} 
at $x$. Now, $\hat L$ is a scalar-valued 4-form constructed from tensorial and 
spinorial quantities and it is therefore invariant under frame transformations, 
whereas $L$ is likewise invariant, by hypothesis. Hence (\ref{normL})
holds for all frames at $x$. The same argument holds at every event $x$ and 
consequently the result (\ref{K}) is proved.

We are now in a position to study the consequences of the various symmetries
of the action. Independent variations of the arguments yield for the matter
Lagrangian
\begin{eqnarray}
\delta L &=&\delta g_{\alpha\beta}\,{\frac {\partial L} {\partial 
g_{\alpha\beta}}} +
\delta\vartheta^\alpha\wedge{\frac {\partial L}{\partial\vartheta^\alpha}} + 
\delta T^\alpha\wedge{\frac {\partial L}{\partial T^\alpha}} +
\delta R_{\alpha}{}^{\beta}\wedge{\frac {\partial L} {\partial
R_{\alpha}{}^{\beta}}}\nonumber\\
 &+& \delta\Psi^A\wedge{\frac {\partial L} {\partial\Psi^A}} +  
\delta(D\Psi^A)\wedge{\frac {\partial L} {\partial (D\Psi^A)}}\,,\label{vard0}
\end{eqnarray}
where the partial derivatives are implicitly {\it defined} by (\ref{vard0}).
Note that in order to avoid counting the nondiagonal components twice in the 
variation procedure, a {\it strict ordering} of the indices is assumed 
in the first term of (\ref{vard0}). The variation $\delta$ and the exterior 
derivative $d$ commute, i.e.\ $[\delta\,,\, d]=0$, since from the very 
definition of the variation of a $p$--form $\delta\Psi:=\Psi'-\Psi$ it
follows $d\delta\Psi:=d\Psi'-d\Psi=\delta d\Psi$. Using this fact, 
we can transform the variations with respect to the torsion and curvature
$T^{\alpha}$ and $R_\alpha{}^\beta$ into the variations with respect to the 
gravitational potentials $\vartheta^\alpha$, and $\Gamma_\alpha{}^\beta\,$. 
We find 
\begin{equation}
\delta T^\alpha = D\delta\vartheta^\alpha + 
\delta\Gamma_\beta{}^\alpha\wedge\vartheta^\beta,\quad\quad 
\delta R_\alpha{}^\beta=D\delta\Gamma_\alpha{}^\beta, 
\end{equation}
and thus
\begin{eqnarray}
\delta L &=&{\frac 1 2}\delta g_{\alpha\beta}\,\sigma^{\alpha\beta} +  
\delta\vartheta^{\alpha}\wedge\Sigma_\alpha+
\delta\Gamma_\alpha{}^{\beta}\wedge\tau^\alpha{}_\beta 
+\delta\Psi^A\wedge{\frac{\delta L}{\delta\Psi^A}}\label{vard1}\\
&&\quad + d\left[{\frac 1 2}\delta g_{\alpha\beta}\ \vartheta^\alpha\wedge
{\frac{\partial L}{\partial T_\beta}}+ \delta\vartheta^{\alpha}\wedge{\frac
{\partial L}{\partial T^{\alpha}}}+ \delta\Gamma_{\alpha}{}^{\beta}
\wedge{\frac{\partial L}{\partial R_{\alpha}{}^{\beta}}}+
\delta\Psi^A\wedge{\frac{\partial L}{\partial D\Psi^A}}\right]\, .\nonumber
\end{eqnarray}
Here, for a gauge--invariant Lagrangian $L$, the expression
\begin{equation}
{\frac {\delta L} {\delta\Psi^A}}
={\frac {\partial  L}{\partial\Psi^A}} - (-1)^{p}D\,
{\frac {\partial L}{\partial (D\Psi^A)}} \label{psi0}
\end{equation}
is the covariant {\it variational derivative} of $L$ with respect to the 
matter $p$--form $\Psi^A$. The {\it matter currents} in (\ref{vard1}) are given by
\begin{eqnarray}
\sigma^{\alpha\beta} &:=& 2\,{\frac {\delta L}{\delta g_{\alpha\beta}}} = 2\,
{\frac {\partial L}{\partial g_{\alpha\beta}}} - D\left(\vartheta^{(\alpha}
\wedge{\frac{\partial L}{\partial T_{\beta)}}}\right),\label{sigM0}\\
\Sigma_{\alpha}&:=& {\frac {\delta L}{\delta\vartheta^{\alpha}}} =  
 {\frac {\partial L}{\partial\vartheta^{\alpha}}}+
D\,{\frac {\partial L}{\partial T^{\alpha}}}\, ,\label{sigC0}\\
\tau_{\alpha\beta} &:=& {\frac {\delta L}{\delta\Gamma_{[\alpha\beta]}}} =   
\rho_{\alpha\beta}{}^A_B\,\Psi^B\wedge{\frac {\partial L} {\partial 
(D\Psi^A)}} + \vartheta_{[\alpha}\wedge 
{\frac {\partial L}{\partial T^{\beta]}}} +
D\, {\frac {\partial L}{\partial R_{\alpha\beta}}}\, .\label{spin0}
\end{eqnarray}
The equations (\ref{vard1})-(\ref{spin0}) were derived with an account
of the vanishing nonmetricity (\ref{zeroQ}), from which the variation of
the symmetric part of connection is expressed in terms of the variation
of the metric,
\begin{equation}
\delta\Gamma_{(\alpha\beta)}={\frac 1 2}D\delta g_{\alpha\beta}.
\end{equation}
This ultimately leaves only the antisymmetric part of the connection 
$\Gamma_{[\alpha\beta]}$ as an independent variable. 

The last term on the right hand side of (\ref{vard1}) is an exact form 
which does not contribute to the action integral because of 
the usual assumption that $\delta g_{\alpha\beta}=0$, 
$\delta\vartheta^{\alpha}=0$, $\delta\Gamma_{\alpha}{}^{\beta}=0$, 
and $\delta\Psi=0$ on the boundary $\partial U$ of the spacetime
domain $U$ of integration.

\subsection{Energy--momentum}

The $4$--form $\sigma^{\alpha\beta}$ and the $3$--form $\Sigma_\alpha$  
are the {\it metrical} (Hilbert) and the {\it ca\-non\-i\-cal} (Noether) 
{\it energy--momentum currents}, respectively.

Since the metric $g_{\alpha\beta}$ can be completely gauged away by the
the frame transformations, it is clear that $\sigma^{\alpha\beta}$ is a 
secondary object, the very existence of which is due to the arbitrariness
of the choice of the frames. That conclusion will be clarified later:
after we have the Noether theorems at our disposal, we will demonstrate that 
the metrical energy-momentum is related to the symmetric part of the canonical 
energy-momentum. 

On the contrary, the canonical energy--momentum 3--form has a clear 
physical meaning as the Noether current corresponding to the local 
translational (general coordinate) invariance of the field theory. 
It is an important dynamical object in the structure of the gravity
theory.

{}From the canonical energy--momentum current we can extract its {\it trace} 
\begin{equation}
\vartheta^{\alpha}\wedge\Sigma_{\alpha},
\end{equation}
with one independent component, and find 
\begin{equation}
\Sigma\quer_{\alpha}:=\Sigma_{\alpha} - {\frac 1 4}
\,e_{\alpha}\rfloor (\vartheta^{\gamma}\wedge\Sigma_{\gamma})\,,\label{sigC1}
\end{equation}
that is traceless:
\begin{equation}
\vartheta^{\alpha}\wedge\Sigma\quer_{\alpha}=0\, . \label{sigC2}
\end{equation}
The {\it antisymmetric} piece $\vartheta_{[\alpha}\wedge 
\Sigma_{\beta ]}$ is a 4-form which has $6$ independent components, 
exactly as the scalar--valued $2$--form
\begin{equation}
\Sigma :=g^{\alpha\beta} e_{\alpha}\rfloor\Sigma_{\beta}= 
e_{\alpha}\rfloor\Sigma^{\alpha}\, .\label{sigC3}
\end{equation}
With the help of some contractions, we find
\begin{equation}
\vartheta_{[\alpha}\wedge \Sigma_{\beta ]}={\frac 1 2}
\vartheta_{\alpha}\wedge\vartheta_{\beta}\wedge\Sigma\, .\label{sigC4}
\end{equation}
Consequently, the irreducible decomposition of the canonical 
energy--mom\-entum $3$--form $\Sigma_\alpha$ into a symmetric tracefree, trace, 
and antisymmetric piece reads
\begin{equation}
\Sigma_{\alpha}= {\buildrel\frown\over{\Sigma\quer}}_{\alpha} +
{\frac 1 4}\,e_{\alpha}\rfloor (\vartheta^{\gamma}\wedge\Sigma_{\gamma}) + 
{\frac 1 2}\,\vartheta_{\alpha}\wedge \Sigma\, .\label{sigC5}
\end{equation}
This equation can be understood as defining the symmetric 
tracefree piece ${\buildrel\frown\over{\Sigma\quer}}_{\alpha}$ with its 
$9$ components. For the symmetric piece
\begin{equation} 
{\buildrel\frown\over{\Sigma}}_{\alpha} =\Sigma_\alpha -{\frac 1 2}
\,\vartheta_\alpha\wedge\Sigma\,,\label{sigC6}
\end{equation}
we find
\begin{equation}
e_{\alpha}\rfloor\,{\buildrel\frown\over{\Sigma}}{}^\alpha=0\, ,\qquad  
\vartheta^{\alpha}\wedge {\buildrel\frown\over{\Sigma\quer}}_{\alpha}=0\,,
\qquad{\rm and}\qquad e_{\alpha}\rfloor\, {\buildrel\frown\over
{\Sigma\quer}}{}^{\alpha}=0\, .\label{sigC7}
\end{equation} 
Moreover, in analogy to (\ref{sigC4}), we have
\begin{equation} 
\vartheta_{(\alpha}\wedge \Sigma_{\beta )}=\vartheta_{(\alpha}\wedge  
{\buildrel\frown\over{\Sigma\quer}}_{\beta )}+ {\frac 1 4}\, 
g_{\alpha\beta}\,(\vartheta^{\gamma}\wedge\Sigma_{\gamma})\, .
\end{equation}

\subsection{Spin current}

The (dynamical) {\it spin current} 3--form
\begin{equation} 
\tau_{\alpha\beta}:= \vartheta_{[\alpha}\wedge\mu_{\beta]}\label{taumu}
\end{equation} 
can be equivalently expressed in terms of a vector--valued $2$--form 
$\mu_\a$. As a first step to prove this, observe that the antisymmetric
3-form $\tau_{\alpha\beta}=\tau_{[\alpha\beta]}$ has the same number of
independent components (namely, 24), as $\mu_\a$. Now, let us find the explicit 
form of {\it the spin energy potential} $2$--form $\mu_\a$. Contracting
(\ref{taumu}) with $e^\beta$, we obtain
\begin{eqnarray}
e^\beta\rfloor\tau_{\alpha\beta} &=& {\frac 1 2}\left(\mu_\alpha -
 \vartheta_{\alpha}\wedge (e^\beta\rfloor\mu_\beta) - 4\mu_\alpha 
+ \vartheta_\beta\wedge e^\beta\rfloor\mu_\alpha\right)\nonumber\\
&=& -{\frac 1 2}\left(\mu_\alpha + \vartheta_{\alpha}\wedge 
(e^\beta\rfloor\mu_\beta)\right).\label{taumu1}
\end{eqnarray}
The second contraction with $e^\alpha$ yields
\begin{equation}
e^\alpha\rfloor e^\beta\rfloor\tau_{\alpha\beta}=
-2e^\beta\rfloor\mu_\beta,
\end{equation}
and substituting this into (\ref{taumu1}), we find finally:
\begin{equation}
\mu_\alpha=-2e^\beta\rfloor \tau_{\alpha\beta} + {\frac 1 2}  
\vartheta_\alpha\wedge(e^\beta\rfloor e^\gamma\rfloor\tau_{\beta\gamma})\,\,.
\label{mutau}
\end{equation}

The 3-form spin current can be decomposed with respect to the $\eta_\mu$
basis of the space of 3-forms,
\begin{equation}
\tau_{\alpha\beta}=\tau_{\mu\alpha\beta}\eta^\mu.
\end{equation}
The components $\tau_{\mu\alpha\beta}$ comprise the {\it spin density tensor}.
It is easy to see that 
\begin{equation}
e^\beta\rfloor\tau_{\alpha\beta}=-\tau_{\mu\nu\alpha}\eta^{\mu\nu},\quad
\quad e^\alpha\rfloor e^\beta\rfloor\tau_{\alpha\beta}= - 
\tau_{\mu\nu\alpha}\eta^{\mu\nu\alpha},
\end{equation}
and hence the spin energy potential reads:
\begin{equation}
\mu_\alpha={\frac 1 2}\left(\tau_{\mu\nu\alpha} + \tau_{\nu\alpha\mu} -
\tau_{\alpha\mu\nu}\right)\eta^{\mu\nu},\label{mutau1}
\end{equation}
which realises its expansion with respect to the $\eta$-basis of 2-forms.

The dynamical spin $\tau_{\alpha\beta}$ is an additional source term which
has the equal importance as the energy--momentum current $\Sigma_{\alpha}$
in the Einstein--Cartan theory, and in a broader context, in
the Poincar\'e gauge theory of gravity.

The field equation for the matter fields $\Psi^A$ is given  
by the familiar Euler--Lagrange equation  
\begin{equation} 
{\frac {\delta L}{\delta\Psi^A}}=0.\label{onsh}
\end{equation}  
If (\ref{onsh}) is assumed to be fulfilled in the course of the derivation 
of identities, we call the latter  the {\it weak} identities in the following 
(``on shell'' in the parlance of the particle physicists).

\section{Noether identities for energy-momentum and spin currents}\label{noeth}

According to the Noether theorem, the conservation identities of the
matter system result from the postulated invariance of $L$ under a
local symmetry group. Actually, this is only true ``weakly'', i.e., 
provided the Euler--Lagrange equation (\ref{onsh}) for the matter fields 
is satisfied.

Here we consider the consequences of the invariance of $L$ under the group
of diffeomorphisms on the spacetime manifold $M$, and under the linear 
transformations of the frame field according to (\ref{genlin}).

\subsection{Diffeomorphisms}
 
Let $\xi$ be a vector field generating an arbitrary one--parameter group 
${\cal T}_{t}$ of diffeomorphisms. In order to obtain the {\it covariant}
Noether identity from the invariance of $L$ under the one--parameter group
of {\it local} translations ${\cal T}_{t} \subset {\cal T}\approx {\it
Di\!f\!f}(4,R)$, we need the conventional Lie
derivative $\ell_\xi:=\xi\rfloor d+ d\xi\rfloor$ on $M$ with respect to
$\xi$. Since our Lagrangian $L$ is also assumed to be a {\it scalar} 
under the linear transformations of the frames, we can equivalently 
replace $\ell_\xi$ by the covariant Lie derivative ${\hbox{\L}}_\xi
:=\xi\rfloor D+D\xi\rfloor$. Then we find directly the
{\it covariant} Noether identity by substituting
${\hbox{\L}}_\xi$ into (\ref{vard0}):
\begin{eqnarray}
{\hbox{\L}}_\xi L &=& ({\hbox{\L}}_\xi\,g_{\alpha\beta})\,
{\frac{\partial L}{\partial g_{\alpha\beta}}}+ ({\hbox{\L}}_{\xi}
\vartheta^\alpha)\wedge{\frac{\partial L} {\partial\vartheta^\alpha}}+ 
({\hbox{\L}}_{\xi} T^\alpha)\wedge{\frac{\partial L} {\partial T^\alpha}} +
({\hbox{\L}}_{\xi} R_{\alpha}{}^{\beta})\wedge{\frac{\partial L} 
{\partial R_{\alpha}{}^{\beta}}}\nonumber\\ &&\quad +({\hbox{\L}}_\xi\Psi^A)
\wedge{\frac{\partial L} {\partial\Psi^A}}+ ({\hbox{\L}}_{\xi}D\Psi^A)\wedge 
{\frac{\partial L} {\partial D\Psi^A}}\,.\label{vard2}
\end{eqnarray}
Recall that the interior product $\xi\rfloor$, which formally acts analogously 
to a derivative of degree $-1$, obeys the Leibniz rule. Since the Lagrangian $L$ 
is the $4$--form on a four-dimensional manifold, its Lie derivative reduces to 
${\hbox{\L}}_\xi L=D(\xi\rfloor L)$. Similarly, since the metric $g_{\alpha\beta}$ 
is a $0$-form with the vanishing covariant derivative, we have ${\hbox{\L}}_\xi 
g_{\alpha\beta}=0$. After expanding the Lie derivatives and performing some 
rearrangements, we get 
\begin{eqnarray}
D(\xi\rfloor L) &=& D\Bigl[(\xi\rfloor\vartheta^{\alpha})\, 
{\frac{\partial L}{\partial\vartheta^{\alpha}}}+ (\xi\rfloor T^{\alpha})\wedge 
{\frac{\partial L}{\partial T^\alpha}}+(\xi\rfloor R_{\alpha}{}^{\beta})\wedge 
{\frac{\partial L}{\partial R_{\alpha}{}^{\beta}}}\nonumber\\ 
&& + (\xi\rfloor\Psi^A)\wedge{\frac{\partial L}{\partial\Psi^A}}+
(\xi\rfloor D\Psi^A)\wedge{\frac{\partial L}{\partial D\Psi^A}}\Bigr]
\nonumber\\ &&- (\xi\rfloor\vartheta^\alpha)D
  {\frac{\partial L}{\partial\vartheta^{\alpha}}}+
(\xi\rfloor T^\alpha)\wedge
  {\frac{\partial L}{\partial\vartheta^{\alpha}}}+
(\xi\rfloor\vartheta^{\alpha})\, R_{\alpha}{}^{\beta}\wedge
  {\frac{\partial L}{\partial T^{\beta}}}\nonumber\\
&&+(\xi\rfloor T^\alpha)\wedge D{\frac{\partial L}{\partial T^{\alpha}}}
+ (\xi\rfloor R_\beta{}^{\gamma})\wedge \vartheta^{\beta}\wedge 
{\frac{\partial L}{\partial T^{\gamma}}}+
(\xi\rfloor R_\beta{}^{\gamma})\wedge 
D{\frac{\partial L}{\partial R_{\beta}{}^{\gamma}}}\nonumber\\
&&+(\xi\rfloor R_\beta{}^\gamma)\wedge\,\rho^{\beta A}_{\ \gamma B}\Psi^B
\wedge{\frac{\partial L}{\partial D\Psi^A}}\nonumber\\
&& +(\xi\rfloor D\Psi^A)\,\wedge{\frac{\delta L}{\delta\Psi^A}}
+(-1)^p(\xi\rfloor\Psi^A)\wedge D{\frac{\delta L}{\delta\Psi^A}}.\label{vard3}
\end{eqnarray}

Collecting together the terms which form the variational derivatives, we obtain
\begin{equation}
A+dB=0\,, \label{A+B}
\end{equation}
where
\begin{eqnarray}
A &:=& -(\xi\rfloor\vartheta^{\alpha})D
{\frac{\delta L}{\delta\vartheta^{\alpha}}}+(\xi\rfloor T^{\alpha})
\wedge{\frac{\delta L}{\delta\vartheta^{\alpha}}}+
(\xi\rfloor R_\beta{}^{\gamma})\wedge
{\frac{\delta L}{\delta \Gamma_{\beta}{}^{\gamma}}}\nonumber\\
&& +(\xi\rfloor D\Psi^A)\,\wedge{\frac{\delta L}{{\delta\Psi^A}}}
  +(-1)^{p}(\xi\rfloor\Psi^A)\wedge D{\frac{\delta L}{\delta\Psi^A}}\, ,\\
B &:=&\xi\rfloor L-\Bigl[(\xi\rfloor\vartheta^\alpha)\,
{\frac{\partial L}{\partial\vartheta^{\alpha}}}+
(\xi\rfloor T^\alpha)\wedge{\frac{\partial L}{\partial T^\alpha}}+
(\xi\rfloor R_{\alpha}{}^{\beta})\wedge
{\frac{\partial L}{\partial R_{\alpha}{}^{\beta}}}\nonumber\\
&& + (\xi\rfloor\Psi^A)\wedge{\frac{\partial L}{\partial\Psi^A}}+
(\xi\rfloor D\Psi^A)\wedge{\frac{\partial L}{\partial D\Psi^A}}\Bigr]\,.
\end{eqnarray}
The functions $A$ and $B$ have the form
\begin{equation}
A=\xi^\a\,A_\a,\qquad\quad B=\xi^\a\,B_\a\,.
\end{equation}
Thus (\ref{A+B}) yields
\begin{equation}
\xi^\a(A_\a+dB_\a)+d\xi^\a\wedge B_\a=0,
\end{equation}
where {\it both} $\xi^\a$ {\it and\/} $d\xi^\a$ are {\it pointwise 
arbitrary}. Hence we conclude that both $B_\a$ and $A_\a$ vanish 
\begin{equation}
A=0,\quad {\rm and}\quad B=0\, .
\end{equation}

{}From $B=0$ we can read off the identity
\begin{eqnarray}
\xi\rfloor L &=&(\xi\rfloor\vartheta^\alpha)\,
{\frac{\partial L}{\partial\vartheta^{\alpha}}}+
(\xi\rfloor T^\alpha)\wedge{\frac{\partial L}{\partial T^\alpha}}+
(\xi\rfloor R_{\alpha}{}^{\beta})\wedge
{\frac{\partial L}{\partial R_{\alpha}{}^{\beta}}}\nonumber\\
&& + (\xi\rfloor\Psi^A)\wedge{\frac{\partial L}{\partial\Psi^A}}+
(\xi\rfloor D\Psi^A)\wedge{\frac{\partial L}{\partial D\Psi^A}}\,.\label{B0}
\end{eqnarray}
After replacing the vector field by the vector basis, 
$\xi\rightarrow e_\alpha$, Eq.(\ref{B0}) yields directly the
explicit form of the {\it canonical energy--momentum current}
\begin{eqnarray}
\Sigma_\alpha &=& e_\alpha\rfloor L-(e_\alpha\rfloor D\Psi^A)\wedge
  {\frac{\partial L}{\partial D\Psi^A}}-(e_\alpha\rfloor\Psi^A)\wedge
  {\frac{\partial L}{\partial\Psi^A}}\nonumber\\
&&+ D{\frac{\partial L}{\partial T^\alpha}}-
(e_{\alpha}\rfloor T^\beta)\wedge{\frac{\partial L}{\partial T^\beta}} 
- (e_{\alpha}\rfloor R_{\beta}{}^{\gamma})\wedge 
{\frac{\partial L}{\partial R_{\beta}{}^{\gamma}}}.\label{momC}
\end{eqnarray}
The first line in (\ref{momC}) represents the result known in the context
of the {\it special} relativistic classical field theory. For the case of
the Maxwell electrodynamics, for example, $\Psi$ stands for the
electromagnetic potential one--form $A=A_i\, dx^i$, with the field
strength two--form $F=DA=dA$. Then (\ref{momC}) describes 
Minkowski's $U(1)$--gauge invariant canonical energy--momentum current
of the Maxwell field. The second line in (\ref{momC}) accounts for the 
possible Pauli terms as well as for the Lagrange multiplier terms in 
the variations with the constraints and it is absent for the
case of the minimal coupling.

{}From $A=0$, we read off the {\it first Noether identity}
\begin{eqnarray}
D\Sigma_\alpha &\equiv & (e_\alpha\rfloor T^\beta)\wedge\Sigma_\beta+
(e_\alpha\rfloor R_{\beta}{}^{\gamma})\wedge\tau^{\beta}{}_{\gamma}
+\,W_{\alpha}\nonumber\\ 
& \cong & (e_\alpha\rfloor T^\beta)\wedge\Sigma_\beta+
(e_\alpha\rfloor R_{\beta}{}^{\gamma})\wedge\tau^{\beta}{}_{\gamma}
, \qquad {\rm (1st)}\label{conmomC}
\end{eqnarray}
where
\begin{equation}
W_{\alpha}:=(e_\alpha\rfloor D\Psi^A)\,{\frac{\delta L}{\delta\Psi^A}}
+(-1)^p(e_\alpha\rfloor\Psi^A)\wedge D{\frac{\delta L}{\delta\Psi^A}}\,. 
\end{equation}
The first line in (\ref{conmomC}) is given in the {\it strong} form, 
without using the field equations. Note that the metrical energy-momentum
$\sigma^{\alpha\beta}$ does not show up in the conservation law at all
which proves its non-dynamical character. 

In the right hand side of the differential identity (\ref{conmomC}) for 
the canonical energy--momentum current we find the typical Lorentz--type 
force terms. They have the general structure {\it field strength}
$\times$ {\it current}.

\subsection{Lorentz invariance and general frame transformation}
 
The invariance of $L$ with respect to the local Lorentz transformations
(\ref{so13}) of the frames gives rise to a further identity. Under 
the infinitesimal transformations (\ref{lor0})-(\ref{lor1}), the  
variations of the geometrical objects and of the matter fields read
as follows  
\begin{equation}
\delta g_{\alpha\beta} =  0,\quad \delta\vartheta^{\alpha} = -  
\omega_{\beta}{}^{\alpha}\,\vartheta^{\beta},\quad
\delta\Gamma_{\alpha}{}^{\beta} = 
D\omega_{\alpha}{}^{\beta},\quad\delta\Psi^A =- 
\omega^{\alpha\beta}\,\rho_{\alpha\beta}{}^A_B\,\Psi^B.\label{lor2}
\end{equation}  
If we insert (\ref{lor2}) into (\ref{vard1}), we obtain
\begin{eqnarray}
\delta L &=&-\omega^{\alpha\beta}\,\Bigl(
\vartheta_{[\alpha}\wedge\Sigma_{\beta]} +D\tau_{\alpha\beta} +  
\rho_{\alpha\beta}{}^A_B\Psi^B\wedge {\frac{\delta L}{\delta\Psi^A}}\Bigr)
\nonumber\\
&& + d\left[\omega^{\alpha\beta}\left(\tau_{\alpha\beta}-
\rho_{\alpha\beta}{}^A_B\Psi^B\wedge{\frac {\partial L}{\partial D\Psi^A}}-
\vartheta_{[\alpha}\wedge{\frac{\partial L}{\partial T^{\beta]}}} -D\,
{\frac{\partial L}{\partial R^{\alpha\beta}}}\right)\right].\label{invL}
\end{eqnarray}
The boundary term vanishes identically in view of the definition
(\ref{spin0}) of the spin current $\tau_{\alpha\beta}$. Then, from the 
arbitrariness of $\omega_\a{}^\b$, we find the {\it second Noether 
identity}  
\begin{equation}
D\tau_{\alpha\beta} +  \vartheta_{[\alpha}\wedge\Sigma_{\beta]}\equiv 
- \rho_{\alpha\beta}{}^A_B\Psi^B\wedge{\frac{\delta L}{\delta\Psi^A}}
\cong 0\,.\qquad {\rm (2nd)}\label{Noe2}
\end{equation}
Again, the weak Noether identity holds provided the matter field
equation (\ref{onsh}) is satisfied.

Now, let us consider the linear transformations of the frame (\ref{genlin})
which are necessarily non-Lorentz. In the infinitesimal form,
\begin{equation}
\Lambda(x)_\alpha{}^\beta = \delta_\alpha^\beta + 
\omega_\alpha{}^\beta,\label{Nlor0}
\end{equation}
but this time, unlike (\ref{lor1}), the transformation parameters are
symmetric
\begin{equation}
\omega_{\alpha\beta} = \omega_{\beta\alpha}.\label{Nlor1}
\end{equation}
Under (\ref{Nlor0})-(\ref{Nlor1}), the geometrical objects and the matter 
fields transform as
\begin{equation}
\delta g_{\alpha\beta} = 2\omega_{\alpha\beta},\quad 
\delta\vartheta^{\alpha} = - \omega_{\beta}{}^{\alpha}\,\vartheta^{\beta},
\quad\delta\Gamma_{\alpha}{}^{\beta} = 
D\omega_{\alpha}{}^{\beta},\quad \delta\Psi^A = 0.\label{Nlor2}
\end{equation}  
Substituting (\ref{Nlor2}) into (\ref{vard1}), we have
\begin{equation}
\delta L = \omega^{\alpha\beta}\left(\sigma_{\alpha\beta} -
\vartheta_{(\alpha}\wedge\Sigma_{\beta)}\right).
\end{equation}
{}From this we find that the metrical energy-momentum is equal to
the symmetric part of the canonical energy-momentum,
\begin{equation}
\sigma_{\alpha\beta}\equiv\vartheta_{(\alpha}\wedge\Sigma_{\beta)}.\label{mC}
\end{equation}
Note that this relation is a {\it strong identity} which holds true in
any Lorentz-covariant field theory when the conditions (\ref{Nlor2}) are 
satisfied. In particular, this is true for the Dirac field (described by the
spinor-valued 0-form) and for the Rarita-Schwinger field (the spinor-valued 
1-form). [The condition (\ref{Nlor2}) is, however, not true in the Proca 
theory where the spin 1 particle is described by the covector-valued 1-form.] 
Thus we see that indeed the metrical energy-mom\-ent\-um current is a secondary 
object which arises as a symmetric part of the canonical energy-momentum current. 
In addition, no conservation law can be established for $\sigma_{\alpha\beta}$ 
directly from the invariance of the Lagrangian under the general coordinate
or the frame transformation. 

In order to get some more insight into the relation between the metrical and
canonical energy-momentum currents, let us introduce instead of the 4-form
$\sigma_{\alpha\beta}$ an equivalent vector-valued 3-form
\begin{equation}
\sigma_\alpha:=e_\beta\rfloor\sigma_\alpha{}^\beta.\label{sigMa}
\end{equation}
Evidently, 
\begin{equation}
e_\beta\rfloor\sigma^\beta=0.
\end{equation}
The identity (\ref{Noe2}) yields for the antisymmetric part of the 
canonical energy-momentum
\begin{equation}
\Sigma=e_\alpha\rfloor\Sigma^\alpha=e_\alpha\rfloor e_\beta\rfloor
D\tau^{\alpha\beta},
\end{equation}
and we then straightforwardly see that the Noether identity (\ref{mC}) 
can be rewritten as
\begin{eqnarray}
\sigma_\alpha &=&{\buildrel\frown\over{\Sigma}}_{\alpha} =
\Sigma_\a-{\frac 1 2}\,\vta_\a\wedge\Sigma \nonumber\\
&=& \Sigma_\alpha - e^\beta\rfloor D\tau_{\alpha\beta}.\label{metcan}
\end{eqnarray}
As it is well known, the relation between the metrical (``Hilbert'') and 
the canonical (``Noether'') energy--momentum currents is established in
the so-called Belinfante-Rosenfeld symmetrization procedure. The last 
formula does not substitute the Belinfante-Rosenfeld result, it rather
can be understood as a specific symmetrization of an otherwise asymmetric 
energy--momentum current for the models which satisfy the condition
(\ref{Nlor2}). 

\section{Gravitational field momenta and Noether identities for the
gravitational Lagrangian}\label{gravlag}

The  total Lagrangian $L_{tot}$ (\ref{lagr0}) includes the pure gravitational
Lagrangian $V$. We assume that the $4$--form $V$ depends on the metric 
$g_{\alpha\beta}$, and the  gravitational potentials $\vartheta^{\alpha},\> 
\Gamma_{\alpha}{}^{\beta}$  and their first derivatives, $d\vartheta^\alpha$ 
and $d\Gamma_\alpha{}^\beta$. By an argument similar to the one used in 
Sec.~\ref{matlag}, we can verify that invariance of $V$ under the tetrad  
deformations requires $V$ to be of the form 
\begin{equation}
V =V(g_{\alpha\beta},\vartheta^{\alpha},T^{\alpha},R_{\alpha}{}^{\beta})\>.
\label{lagrV}
\end{equation}

Consequently, we can use the results of Sec.~\ref{noeth} and apply them 
to the gravitational Lagrangian simply by replacing $L$ by $V$ and by dropping 
all $\Psi$--dependent terms in the end. For convenience, we
condense our notation and introduce, according to the conventional 
canonical prescription, the following {\it gauge field momenta} 2-forms:
\begin{equation} 
H_{\alpha} :=  
- {\frac{\partial V}{\partial T^{\alpha}}}\,,\qquad  
H^{\alpha}{}_{\beta} := 
- {\frac{\partial V}{\partial R_{\alpha}{}^{\beta}}}\, .\label{HH}
\end{equation}  
Moreover, we define the {\it metrical} energy--momentum $4$--form  
\begin{equation} 
m^{\alpha\beta} := 2\,{\frac{\partial V}{\partial g_{\alpha\beta}}},\label{ma}
\end{equation}  
the {\it canonical} energy--momentum $3$--form  
\begin{equation} 
E_{\alpha} := {\frac{\partial V}{\partial\vartheta^{\alpha}}}\label{Ea0}
\end{equation}  
and the {\it spin} $3$--form  
\begin{equation} 
E^{\alpha\beta}:= {\frac{\partial V}{\partial\Gamma_{[\alpha\beta]}}}= -
\vartheta^{[\alpha}\wedge H^{\beta]} \label{Eab}
\end{equation}  
for the gravitational gauge fields themselves. If we  apply the 
variational principle (\ref{vard1}) with respect to the {\it independent 
variables} $g_{\alpha\beta},\, \vartheta^{\alpha}$, and  
$\Gamma_{\alpha}{}^{\beta}$ and compare it with (\ref{sigM0})-(\ref{spin0}), 
we find
\begin{eqnarray}
2\,{\frac{\delta V}{\delta g_{\alpha\beta}}} &=& - DM^{\alpha\beta} + 
m^{\alpha\beta}, \label{dVg}\\  
{\frac{\delta V}{\delta\vartheta^{\alpha}}} &=& - DH_{\alpha}  
+E_{\alpha}, \label{dVt}\\ 
{\frac{\delta V}{\delta\Gamma_{[\alpha\beta]}}} &=& -  
DH^{\alpha\beta} + E^{\alpha\beta}\,.\label{dVG}
\end{eqnarray}
Here
\begin{equation}
M^{\alpha\beta}:=-\vartheta^{(\alpha}\wedge H^{\beta)}\label{Mab}
\end{equation}
plays the role of the metric field momentum.

The Noether machinery can be applied to the gravitational 
Lagrangian (\ref{lagrV}) in a precisely the same way as it was done 
for the material Lagrangian in Sec.~\ref{noeth}.  As a result, we find:

\begin{description} 

\item[(i)] The diffeomorphism invariance yields the explicit structure of the 
canonical energy--momentum $3$--form  
\begin{equation} 
E_{\alpha} = e_{\alpha}\rfloor V + (e_{\alpha}\rfloor  
T^{\beta})\wedge H_{\beta} + (e_{\alpha}\rfloor
R_{\beta}{}^{\gamma})\wedge H^{\beta}{}_{\gamma} \label{Ea}
\end{equation} 
of the gauge fields, cf.\ (\ref{momC}) for the material case.
This implies for its trace 
\begin{equation}
\vartheta^{\alpha}\wedge E_{\alpha} = 4V + 2  
T^{\beta}\wedge H_{\beta} + 2R_{\beta}{}^{\gamma}\wedge H^{\beta}{}_{\gamma}.
\end{equation} 
Furthermore we find the {\it first Noether identity}  
\begin{equation}
D\,{\frac{\delta V}{\delta\vartheta^{\alpha}}}\equiv (e_{\alpha}\rfloor  
T^{\beta})\wedge{\frac{\delta V}{\delta\vartheta^{\beta}}} +  
(e_{\alpha}\rfloor R_{\beta}{}^{\gamma})\wedge\,{\frac{\delta V}  
{\delta\Gamma_{\beta}{}^{\gamma}}},\qquad {\rm (1st)}
\end{equation} 
as a gravitational counterpart of the identity (\ref{conmomC}) for the  
matter Lagrangian.  
 
\item[(ii)] The invariance with respect to the (infinitesimal) local Lorentz
transformations yields the {\it second Noether identity}
\begin{equation}
D\,{\frac{\delta V}{\delta\Gamma^{[\alpha\beta]}}} +  
\vartheta_{[\alpha}\wedge{\frac{\delta V}{\delta\vartheta^{\beta]}}}
\equiv 0. \qquad {\rm (2nd)}\label{NoeV2}
\end{equation} 

\item[(iii)] The invariance with respect to the (infinitesimal) local non-Lorentz
transformations of the frames yields an additional identity
\begin{equation}
2\,{\frac{\delta V}{\delta g_{\alpha\beta}}} - 
\vartheta_{(\alpha}\wedge{\frac{\delta V}{\delta\vartheta^{\beta)}}}
\equiv 0.\label{symN}
\end{equation}
\end{description}

Observe that the Noether identities for the gravitational gauge fields
are all {\it strong}, since no field equation is involved 
in their derivation.

By inserting (\ref{dVg})-(\ref{dVt}) into the Noether identity (\ref{symN}), 
we obtain the following explicit form of the metrical gravitational
energy--momentum current
\begin{equation}
m^{\alpha\beta}= \vartheta^{(\alpha}\wedge E^{\beta)} -
T^{(\alpha}\wedge H^{\beta)} .\label{mab}
\end{equation} 
Consequently, its trace is given by
\begin{equation} 
m^{\alpha}{}_\a =\vartheta^\alpha\wedge E_\alpha 
-T^\alpha\wedge H_\alpha = 4V + T^{\beta}\wedge H_{\beta} +
2R_{\beta}{}^{\gamma}\wedge H^{\beta}{}_{\gamma} \,. 
\end{equation}
Analogously, after using (\ref{dVg})-(\ref{dVt}) and (\ref{Eab}) in 
(\ref{NoeV2}), we find
\begin{equation}
\vartheta_{[\alpha}\wedge E_{\beta]} -
T_{[\alpha}\wedge H_{\beta]} + R_{\alpha}{}^{\gamma}\wedge H_{\gamma\beta}
+ R_{\beta}{}^{\gamma}\wedge H_{\alpha\gamma}=0,\label{NoeV2a}
\end{equation}
where we used the definition of the curvature as a commutator of 
covariant derivatives, $DDH_{\alpha\beta}\equiv R_{\alpha}{}^{\gamma}\wedge 
H_{\gamma\beta} + R_{\beta}{}^{\gamma}\wedge H_{\alpha\gamma}$. 
Alternatively, we may collect the symmetric and skew-symmetric identities
(\ref{mab}) and (\ref{NoeV2a}) into a single equation,
\begin{equation}
m_{\alpha\beta}= \vartheta_{\alpha}\wedge E_{\beta} -
T_{\alpha}\wedge H_{\beta} + R_{\alpha}{}^{\gamma}\wedge H_{\gamma\beta}
+ R_{\beta}{}^{\gamma}\wedge H_{\alpha\gamma}.\label{mab1}
\end{equation} 

\section{Gravitational field equations}\label{fieldeqs}

 Now we are in the position to formulate the action principle in full
generality: The total action of the gravitational gauge fields 
and of the minimally coupled matter fields reads  
\begin{equation}
W = \int [V(g_{\alpha\beta}, \vartheta^{\alpha}, 
T^{\alpha}, R_{\alpha}{}^{\beta}) +  
L(g_{\alpha\beta}, \vartheta^{\alpha},\Psi , D\Psi)].
\end{equation}  
The {\it independent variables} are  $\Psi,\,  g_{\alpha\beta},\,  
\vartheta^{\alpha}$, and $\Gamma_{\alpha}{}^{\beta}$. Their independent 
variation then yields, by means of (\ref{dVg})-(\ref{dVG}) and  
the definitions (\ref{HH})-(\ref{Eab}),(\ref{Mab}) and 
(\ref{psi0})-(\ref{spin0}), the {\it Yang--Mills type gauge field equations} 
of gravity: 
\begin{eqnarray}
{\frac{\delta L}{\delta\Psi}} &=& 0\,,\quad\qquad\qquad\,
{\rm (MATTER)}\label{Pmat}\\ 
DM^{\alpha\beta} - m^{\alpha\beta} &=&  
\sigma^{\alpha\beta}\,, \qquad\,\qquad {\rm (ZEROTH)}\label{Peq0}\\
DH_{\alpha} - E_{\alpha} &=& 
\Sigma_{\alpha}\,, \qquad\>\;\qquad {\rm (FIRST)}\label{Peq1}\\
DH_{\alpha\beta} - E_{\alpha\beta} &=&  
\tau_{\alpha\beta}\,. \qquad\qquad\, {\rm (SECOND)}\label{Peq2}
\end{eqnarray} 
The covariant exterior derivatives $D$ of the gauge
field momenta describe the terms of the Yang--Mills type.
In addition, due to the universality of the gravitational
interaction, we find the {\it self--coupling} terms which
involve the metrical energy--momentum $m^{\a\b}$, the
canonical energy--momentum $E_\a$, or the spin $E_{\a\b}$ of
the gravitational fields, respectively. They, together with the corresponding
material currents $\sigma^{\a\b}$, $\Sigma_\a$, and $\tau_{\a\b}$,
act as sources of the gauge field potentials.

This dynamical framework is very general. As special cases, it contains 
the field equations of the general relativity (GR) theory and those of the 
Einstein--Cartan theory. Both are the dynamically degenerate cases of 
the Poincar\'e gauge theory.

As soon as an explicit gauge Lagrangian $V$ is specified, all we have
to do is {\it to partially differentiate} this Lagrangian with respect
to the field strengths, the torsion  $T^\a$ and the curvature $R_\a{}^\b$,
respectively.  Thereby we find the gauge field momenta in (\ref{HH}).
If we substitute the latter into (\ref{Eab}), (\ref{Ea}), and (\ref{mab})
and, subsequently, into the field equations (\ref{Peq0})-(\ref{Peq2}), 
we obtain the field equations explicitly. Our framework allows to 
investigate the different gauge Lagrangians in a straightforward way. 

Note, in particular, that we do not need to vary the Hodge star,
a computation which would complicate things appreciably. The idea to use
the gauge field momenta as the operationally meaningful quantities in their 
own right --- together with the temporary suspension of the relations
between the momenta and the field strengths --- is taken from the 
Kottler--Cartan--van Dantzig representation of the electrodynamics 
(see the book \cite{Birkbook}).

Compared to the earlier work on this subject, 
in which only the two field equations occur, we have obtained a system 
of the {\it three} gauge field equations for the gravitational potentials.
This can be traced back to the assumption that the coframe field 
$\vartheta^{\alpha}$ is {\it not} necessarily {\it orthonormal}. This 
allows for a more flexibility in the process of the eventual solving of 
the field equations. However, it is clear that the price for this 
flexibility is a certain duplication of the dynamical equations. It is 
straightforward to verify that not all of the gravitational field 
equations (\ref{Peq0})-(\ref{Peq2}) are independent. 

This fact is obvious already from the direct counting of the number of the
field variables. For the description of the gravitational field, we have
$10+16+24=50$ components of the metric $g_{\alpha\beta}$, coframe 
$\vartheta^\alpha$ and connection $\Gamma_{[\alpha\beta]}$. We count only
the antisymmetric part, since the symmetric piece of connection is 
expressed in terms of the metric via (\ref{Gsym}). Formally, we have
exactly $50$ field equations (\ref{Peq0})-(\ref{Peq2}). However, there is
a wide symmetry group of the action which includes the local Lorentz
rotations plus the general linear transformations of the frame (\ref{genlin}).
Together, they amount to $6+10=16$ arbitrary functions which parametrize
these transformations. In principle, we always have an option to ``gauge away''
completely either the metric or the coframe everywhere on the spacetime 
manifold $M$, and to work in one of the following gauges:
\begin{itemize}
\item The {\it constant-metric-gauge} is obtained by choosing the frames in
such a way that the metric
\begin{equation}
g_{\alpha\beta}={\rm constant}\label{metgaug}
\end{equation}
everywhere. By imposing this gauge, we fix the freedom of the non-Lorentz
linear transformations (\ref{genlin}), and reduce the number of gravity
field variables to $16+24=40$ (for the coframe $\vartheta^\alpha$ and 
connection $\Gamma_{[\alpha\beta]}$). Note, however, that the local Lorentz
rotations which, by definition (\ref{so13}), preserve (\ref{metgaug}), are
still available. This remaining freedom (involving 6 arbitrary functions)
can be used for the further reduction of the number of variables. In 
particular, one can eliminate any 6 of the 16 components of the coframe,
so that finally we end up with $10+24=34$ independent variables. For example,
a convenient choice is to make a $4\times 4$ matrix of the coframe coefficients,
$e^\alpha_i$, {\it symmetric}. 

\item {\it Constant-coframe-gauge} is achieved after the combined use of the
local Lorentz and general linear transformations, reducing the coframe to
\begin{equation}
\vartheta^\alpha=\delta^\alpha_i\,dx^i=dx^\alpha\label{congaug}
\end{equation}
everywhere. This eliminates the coframe components completely, and one is
left again with $10+24=34$ independent variables (this time they are the
metric $g_{\alpha\beta}$ and the connection $\Gamma_{[\alpha\beta]}$). One
can call this a holonomic gauge since effectively the holonomic components
of the metric $g_{ij}$ and the metric-compatible connection $\Gamma_i{}^j$
describe the gravitational field configurations now.
\end{itemize}

The possibility of ``gauging away'' either the metric or the coframe 
(\ref{metgaug}), (\ref{congaug}), clearly suggests that exactly $16$ of the
$50$ field equations (\ref{Peq0})-(\ref{Peq2}) are redundant. Indeed, let us
rewrite the first field equation (\ref{Peq1}) in the equivalent form,
\begin{equation} 
\vartheta_{\alpha}\wedge DH_{\beta} - 
\vartheta_{\alpha}\wedge E_{\beta}  = 
\vartheta_{\alpha}\wedge \Sigma_{\beta}\,.
\end{equation}
Substituting the second Noether identity (\ref{NoeV2a}), one can transform 
the left-hand side to 
\begin{eqnarray}
\vartheta_{\alpha}\wedge DH_{\beta} &-& 
T_{\alpha}\wedge H_{\beta} + R_{\alpha}{}^{\gamma}\wedge H_{\gamma\beta}
+ R_{\beta}{}^{\gamma}\wedge H_{\alpha\gamma}- m_{\alpha\beta}\nonumber\\
&=& -D(\vartheta_{\alpha}\wedge H_{\beta}) - DDH_{\alpha\beta} -m_{\alpha\beta}.
\end{eqnarray}
Thus the first equation is equivalent to 
\begin{equation}
-D(\vartheta_{\alpha}\wedge H_{\beta} + DH_{\alpha\beta}) -m_{\alpha\beta}=
\vartheta_{\alpha}\wedge \Sigma_{\beta}\,.
\end{equation}

We can decompose this equation into the symmetric and antisymmetric parts.
Then we immediately see that the {\it antisymmetric part is identically
vanishing} due to the Noether identity (\ref{Noe2}) and the second field
equation (\ref{Peq2}). On the other hand, the {\it symmetric part is 
identically reproducing} the zeroth equation (\ref{Peq0}), because the metric
energy-momentum is equal to the symmetric part of the canonical energy-momentum, 
(\ref{mC}). Since  the matter field equation (\ref{Pmat}) is a prerequisite 
for the validity of the differential Noether identity, we obtain the important 
result that one of the first two gravitational field equations is ``weakly'' 
{\it redundant}, and the number of truly independent field equations is
indeed $50-16=34$. 

Not unexpectedly, the second field equation does {\it not} follow from the other 
field equations. While working in the constant-metric-gauge (\ref{metgaug}), 
it is convenient to solve the coupled system of the second equation and the 
symmetric part of the first field equation. Analogously, in the constant-frame
gauge (\ref{congaug}), it may be more natural to consider the equivalent set of
the coupled system of the zeroth and the second field equations.

\section{Limiting case: spinless matter}\label{spinless}

The Noether identities (\ref{conmomC}) and (\ref{Noe2}) contain a plenty of
information about the interaction of the spin of the classical matter with the
post-Riemannian geometry of a spacetime. They also allow for a transparent limit 
when the spin vanishes $\tau_{\alpha\beta}=0$. Let us study this in some detail. 

Recall that the metric-compatible connection can be decomposed into
the Riemannian and post-Riemannian parts as
\begin{equation}
\Gamma_\alpha{}^\beta = \widetilde{\Gamma}_\alpha{}^\beta - 
K_\alpha{}^\beta.\label{gagaK}
\end{equation}
Here the tilde denotes the purely Riemannian connection and $K_\alpha{}^\beta$
is the contortion which is related to the torsion via the identity
\begin{equation}
T^\alpha=K^\alpha{}_\beta\wedge\vartheta^\beta.\label{contor}
\end{equation}
Accordingly, the curvature is decomposed as
\begin{equation}\label{RK}
R_\beta{}^\gamma = \widetilde{R}_\beta{}^\gamma - 
\widetilde{D}K_\beta{}^\gamma - K_\beta{}^\lambda\wedge K_\lambda{}^\gamma.
\end{equation}
Hence, the last term in the 1st Noether identity (\ref{conmomC}) reads
\begin{eqnarray}
(e_\alpha\rfloor R_\beta{}^\gamma)\wedge\tau^\beta{}_\gamma &=&\,
(e_\alpha\rfloor \widetilde{R}_{\beta\gamma})\wedge\tau^{\beta\gamma}
\nonumber\\
&&- (e_\alpha\rfloor \widetilde{D}K_{\beta\gamma})\wedge\tau^{\beta\gamma} -
e_\alpha\rfloor (K_\beta{}^\lambda\wedge K_{\lambda\gamma})
\wedge\tau^{\beta\gamma}.\label{dec1}
\end{eqnarray}
Using (\ref{contor}), we find
\begin{equation}
e_\alpha\rfloor T^\beta = (e_\alpha\rfloor K^\beta{}_\gamma)\vartheta^\gamma
+ K_\alpha{}^\beta,
\end{equation}
and hence
\begin{eqnarray}
(e_\alpha\rfloor T^\beta)\wedge\Sigma_\beta &=& 
(e_\alpha\rfloor K^{\beta\gamma})\vartheta_{[\gamma}\wedge\Sigma_{\beta]}
+ K_\alpha{}^\beta\wedge\Sigma_\beta\nonumber\\
&=&K_\alpha{}^\beta\wedge\Sigma_\beta + (e_\alpha\rfloor K^{\beta\gamma})
\wedge D\tau_{\beta\gamma}\nonumber\\
&=&K_\alpha{}^\beta\wedge\Sigma_\beta + (e_\alpha\rfloor K^{\beta\gamma})
\wedge\widetilde{D}\tau_{\beta\gamma} + 
e_\alpha\rfloor (K_\beta{}^\lambda\wedge K_{\lambda\gamma})
\wedge\tau^{\beta\gamma},\label{dec2}
\end{eqnarray}
where we used the second Noether identity (\ref{Noe2}), and then decomposed
the covariant derivative into the Riemannian and post-Riemannian parts
according to (\ref{gagaK}). Finally, again with the help of (\ref{gagaK}),
one finds
\begin{equation}
D\Sigma_\alpha = \widetilde{D}\Sigma_\alpha + 
K_\alpha{}^\beta\wedge\Sigma_\beta .\label{dec3}
\end{equation}

Substituting (\ref{dec1}), (\ref{dec2}), (\ref{dec3}) into the 1st Noether
identity, we finally obtain the conservation law
\begin{equation}
\widetilde{D}\Sigma_\alpha = (e_\alpha\rfloor K_{\beta\gamma})
\wedge\widetilde{D}\tau^{\beta\gamma} - (e_\alpha\rfloor \widetilde{D}
K_{\beta\gamma})\wedge\tau^{\beta\gamma} + (e_\alpha\rfloor 
\widetilde{R}_{\beta\gamma})\wedge\tau^{\beta\gamma}.\label{dec4}
\end{equation}
Making use of the definition of the covariant Lie derivative, 
$\widetilde{\hbox{\L}}_{e_\alpha}=e_\alpha\rfloor\widetilde{D} + 
\widetilde{D}e_\alpha\rfloor$, the last equation can be rewritten in the
equivalent form,
\begin{equation}
\widetilde{D}\left(\Sigma_\alpha -\tau^{\beta\gamma}e_\alpha\rfloor 
K_{\beta\gamma}\right)-\tau^{\beta\gamma}\wedge\widetilde{\hbox{\L}}_{e_\alpha}
K_{\beta\gamma} = (e_\alpha\rfloor\widetilde{R}_{\beta\gamma})
\wedge\tau^{\beta\gamma}.\label{Noe1dec}
\end{equation}

The ``decomposed'' form of the first Noether identity, (\ref{Noe1dec}), shows
that the post-Riemannian geometrical variables (contortion) are coupled
directly to the spin of matter. In particular, when the matter is spinless,
$\tau^{\beta\gamma}=0$, we are left with the purely Riemannian conservation
law (although the geometry is still non-Riemannian!):
\begin{equation}
\widetilde{D}\Sigma_\alpha =\widetilde{D}\sigma_\alpha =0.\label{geod}
\end{equation}
The first equality arises from (\ref{metcan}) which shows that the metric
and canonical energy-momenta are coinciding for spinless case. 

An important physical conclusion is thus that test particles without spin
are always moving along the {\it Riemannian geodesics}, in complete
agreement with the equivalence principle. To put it differently, the
spacetime torsion can only be detected with the help of test matter with
spin. 

On the other hand, it is worthwhile to note that in the absence of torsion,
the equation (\ref{Noe1dec}) displays the standard Mathisson-Papapetrou
force of GR which gives rise to a non-geodesic motion of a test particle
with spin. 

\section{General quadratic models}\label{quadratic}

The general Lagrangian which is at most quadratic in the Poincar\'e
gauge field strengths -- in the torsion and the curvature -- reads
\begin{eqnarray} 
V_{\rm Q}&=&
\frac{1}{2\kappa}\,\left[a_0\,R^{\alpha\beta}\wedge\eta_{\alpha\beta}
-2\lambda\,\eta - T^\alpha\wedge{}^*\!\left(\sum_{I=1}^{3}a_{I}\,^{(I)}
T_\alpha\right)\right]\nonumber\\
&&-\,\frac{1}{2}\,R^{\alpha\beta}\wedge{}^*\!\left(\sum_{I=1}^{6}b_{I}\,
{}^{(I)}R_{\alpha\beta}\right).\label{lagr}
\end{eqnarray}
We use the unit system in which the dimension of the gravitational constant 
is $[\kappa] = \ell^2$ with the unit length $\ell$. The coupling constants
$a_0,a_1,a_2,a_3$ and $b_1,...,b_6$ are {\it dimensionless}, whereas 
$[\lambda] = \ell^{-2}$. These coupling constants determine the particle
contents of the qudratic Poincar\'e gauge models, the corresponding analysis 
can be found in \cite{Kuhfuss,Neville,Sezgin1,Sezgin2,OPZ}. 

The Lagrangian (\ref{lagr}) has the general structure similar to that of the 
Yang--Mills Lagrangian for the gauge theory of internal symmetry group.

In order to be able to compare (\ref{lagr}) to the Lagrangians studied in
the literature, let us rewrite $V_{\rm Q}$ using the tensor language:
\begin{eqnarray}
V_{\rm Q}&=& -\,\frac{1}{2}\,\eta\,\Bigg[{\frac 1\kappa}\left(a_0 R
+ 2\lambda + \alpha_1\,T_{\mu\nu}{}^\alpha\,T^{\mu\nu}{}_\alpha + 
\alpha_2\,T_{\mu\alpha}{}^\mu\,T^{\nu\alpha}{}_\nu + \alpha_3\,T_{\mu\nu}{}^\alpha
\,T_\alpha{}^{\mu\nu}\right)\nonumber\\
&&\qquad  + \beta_1\,R_{\mu\nu\alpha\beta}R^{\mu\nu\alpha\beta} 
+ \beta_2\,R_{\mu\nu\alpha\beta}R^{\mu\alpha\nu\beta} 
+ \beta_3\,R_{\mu\nu\alpha\beta}R^{\alpha\beta\mu\nu}\nonumber\\
&&\qquad  + \beta_4\,{\rm Ric}_{\mu\nu}{\rm Ric}^{\mu\nu} + \beta_5
\,{\rm Ric}_{\mu\nu}{\rm Ric}^{\nu\mu} + \beta_6\,R^2\Bigg].\label{lagrT}
\end{eqnarray}
Using the definitions of the irreducible torsion and curvature parts (see 
Appendix 1), we find the relation between the coupling constants:
\begin{eqnarray}
a_1 &=& 2\alpha_1 - \alpha_3,\quad a_2 = 2\alpha_1 + 3\alpha_2 - \alpha_3,
\quad a_3 = 2\alpha_1 + 2\alpha_3,\label{ccc}\\
b_1 &=& 2\beta_1        + 2\beta_3,\\
b_2 &=& 2\beta_1 +  \beta_2 - 2\beta_3,\\
b_3 &=& 2\beta_1 - 2\beta_2 + 2\beta_3,\\
b_4 &=& 2\beta_1 +  \beta_2 + 2\beta_3 +  \beta_4 +  \beta_5,\\
b_5 &=& 2\beta_1        - 2\beta_3 +  \beta_4 -  \beta_5,\\
b_6 &=& 2\beta_1 +  \beta_2 + 2\beta_3 + 3\beta_4 + 3\beta_5 + 12\beta_6.
\end{eqnarray}
The inverse of (\ref{ccc}) reads
\begin{equation}
\alpha_1 = {\frac {2a_1 + a_3} 6},\quad \alpha_2 = {\frac {a_2 - a_1} 3},
\quad \alpha_3 = {\frac {a_3 - a_1} 3}.
\end{equation}

The Poincar\'e gauge field equations are derived from the total
Lagrangian $V_{\rm Q} + L_{\rm mat}$ within the general framework described
in Sec.~\ref{fieldeqs}. The resulting system of the {\it first} (\ref{Peq1}) 
and {\it second} (\ref{Peq2}) field equations reads 
\begin{eqnarray} 
DH_{\alpha}- E_{\alpha}&=&\Sigma_{\alpha}\,,\label{first}\\ 
DH^{\alpha}{}_{\beta}- E^{\alpha}{}_{\beta}&=&\tau^{\alpha}{}_{\beta}\,.
\label{second}
\end{eqnarray} 
The right hand sides describe the material sources of the Poincar\'e gauge
gravity: the canonical energy--momentum (\ref{sigC0}) and the spin 
(\ref{spin0}) three--forms. 
 
The explicit form for the gauge field momenta (\ref{HH}) which enter the
left hand sides of (\ref{first})--(\ref{second}) is given by
\begin{eqnarray}
H_{\alpha}:=-{\frac {\partial V_{\rm Q}} {\partial T^\alpha}} &=& 
{1\over\kappa}\,{}^*\!\left(\sum_{I=1}^{3}a_{I}{}^{(I)}
T_{\alpha}\right),\label{Ha1}\\  
H^{\alpha}{}_{\beta}:= - {\frac {\partial V_{\rm Q}} {\partial R_{\alpha}
{}^\beta}} &=& -{\frac {a_0} {2\kappa}}\,\eta^{\alpha}{}_{\beta} + {}^{*}\!
\left(\sum_{I=1}^{6}b_{I}\,{}^{(I)}R^{\alpha}{}_{\beta}\right).\label{Hab1}
\end{eqnarray}
The three-forms $E_{\alpha}$ and $E^{\alpha}{}_{\beta}$ describe, 
respectively, the {\it canonical energy-momentum} and {\it spin} densities 
determined by the Poincar\'e gauge gravitational field via (\ref{Ea0}) and
(\ref{Eab}).

\section{Einstein's theory -- a degenerate case of quadratic 
Poincar\'e gravity}\label{Einsteinlimit}

In Einstein's general relativity theory, the metric is the only fundamental
field variable. The linear connection is constructed from the first derivatives
of the metric components, it is a unique metric compatible and torsion-free
connection which we denote hereafter $\widetilde{\Gamma}_\alpha{}^\beta$. In 
the local Lorentz invariant formulation, the metric is effectively replaced 
by the (orthonormal) coframe, $\vartheta^\alpha=e^\alpha_i\,dx^i$, which is 
a kind of a ``square root'' of the spacetime metric: $g_{ij}=e^\alpha_i\,
e^\beta_j\,o_{\alpha\beta}$. In terms of the metric/coframe components, the
purely Riemannian connection reads
\begin{equation}
\widetilde{\Gamma}_{\alpha\beta} = e_{[\alpha}\rfloor C_{\beta]} -{\frac 1 2}
\,(e_\alpha\rfloor e_\beta\rfloor C_\gamma)\,\vartheta^\gamma,\label{christ}
\end{equation}
where the anholonomity two--form is defined as usual,
\begin{equation}
C^\alpha:=d\vartheta^\alpha.\label{anhol}
\end{equation}
It is straightforward to check that the {\it Christoffel symbol} (\ref{christ})
indeed has zero torsion and nonmetricity:
\begin{equation}
\widetilde{D}\vartheta^\alpha\equiv 0,\quad\quad
\widetilde{D}g_{\alpha\beta}\equiv 0.\label{zeroTQ}
\end{equation}
Hereafter the differential operators defined by the Riemannian connection
and geometrical objects constructed from (\ref{christ}) (e.g., the exterior
covariant differential $\widetilde{D}$ and  the curvature
$\widetilde{R}_\alpha{}^\beta$ two--form) will be denoted by the tilde. 
In view of (\ref{zeroTQ}), the Riemannian covariant derivatives of the
dual $\eta$-forms are also vanishing,
\begin{equation}
\widetilde{D}\eta_{\mu_1\dots\mu_p}\equiv 0,\quad\quad
p=0,1,...,4.\label{zeroeta}
\end{equation}

In this section we will demonstrate that the standard general relativity
(Einstein's theory of gravity) is the special case of the Poincar\'e gauge
gravity. It is described by the particular quadratic model (\ref{lagr}) 
when the coupling constants are chosen as follows:
\begin{equation}
a_0=1,\quad a_1=-1,\quad a_2=2,\quad a_3={\frac 1 2},
\quad\quad b_I=0.\label{afixed}
\end{equation}
This choice is {\it degenerate} in the sense we explain below. Let us
write down explicitly the Lagrangian of this model,
\begin{equation}
V^{(0)} = {\frac 1 {2\kappa}}\left(R^{\mu\nu}\wedge\eta_{\mu\nu} 
+ {}^{(1)}T^{\alpha}\wedge{}^{*(1)}T_{\alpha}
- 2{}^{(2)}T^{\alpha}\wedge{}^{*(2)}T_{\alpha} -
{1\over 2}{}^{(3)}T^{\alpha}\wedge{}^{*(3)}T_{\alpha}\right).\label{V0lag}
\end{equation}
In accordance with the general scheme we find the gauge momenta for 
(\ref{V0lag}):
\begin{eqnarray}
H^{(0)}_{\alpha}=-{\frac {\partial V^{(0)}} {\partial T^\alpha}} &=& 
\,{1\over \kappa}\,{}^*\!\left(-{}^{(1)}T_{\alpha} + 2\,{}^{(2)}T_{\alpha} 
+ {1\over 2}\,{}^{(3)}T_{\alpha}\right),\label{Ha0}\\  
H^{(0)\alpha}{}_{\beta}= - {\frac {\partial V^{(0)}} {\partial R_{\alpha}
{}^\beta}} &=& -{\frac 1 {2\kappa}}\,\eta^{\alpha}{}_{\beta}.\label{Hab0}
\end{eqnarray}

The degeneracy of the model under consideration is manifested in the
fact that the left hand side of the {\it second field equation is identically 
zero}:
\begin{equation}
DH^{(0)\alpha}{}_{\beta} - E^{(0)\alpha}{}_{\beta}\equiv 0.\label{secondL}
\end{equation}
Here, see (\ref{Eab}), $E_{\alpha\beta}=-\vartheta_{[\alpha}\wedge 
H^{(0)}_{\beta]}$. The proof relies on two geometrical identities, (\ref{ID1})
and (\ref{ID2}), which we prove in Appendix 3 at the end of the paper.
In particular, of fundamental importance is to observe that
\begin{equation}
H^{(0)}_{\alpha}\equiv 
{\frac 1 {2\kappa}}\,K^{\mu\nu}\wedge\eta_{\alpha\mu\nu}\label{H0K}
\end{equation}
which follows from the identity (\ref{ID1}). 

Let us now turn to the {\it first} field equation. With the help of 
(\ref{Ha0}), one can rewrite the Lagrangian (\ref{V0lag}) as
\begin{equation}
V^{(0)} = {\frac 1 {2\kappa}}\,R^{\mu\nu}\wedge\eta_{\mu\nu} -
 {\frac 1 2}\,T^\beta\wedge H^{(0)}_\beta.
\end{equation}
Hence, see (\ref{Ea}), the gravitational energy--momentum three-form
reads
\begin{eqnarray}
E^{(0)}_\alpha &=& e_{\alpha}\rfloor V^{(0)} + (e_{\alpha}\rfloor T^{\beta})
\wedge H^{(0)}_{\beta} + (e_{\alpha}\rfloor R_{\beta}{}^{\gamma})\wedge
H^{(0)\beta}{}_{\gamma}\nonumber\\
&=& {\frac 1 {2\kappa}}\,R^{\mu\nu}\wedge\eta_{\alpha\mu\nu} + 
{\frac 1 2}\,(e_{\alpha}\rfloor T^\beta)\wedge H^{(0)}_\beta -
{\frac 1 2}\,T^\beta\wedge (e_{\alpha}\rfloor H^{(0)}_\beta),\label{Ea0f}
\end{eqnarray}
where we substituted (\ref{Hab0}).

Some efforts are required to calculate the term  
\begin{equation}
DH^{(0)}_{\alpha}=\widetilde{D}H^{(0)}_{\alpha} - K_\alpha{}^\beta\wedge
H^{(0)}_\beta.\label{DH0}
\end{equation}
At first, directly from the identity (\ref{H0K}) we find:
\begin{equation}
T^\beta\wedge (e_{\alpha}\rfloor H^{(0)}_\beta) = {\frac 1 {2\kappa}}\left(
(e_{\alpha}\rfloor K^{\mu\nu})\,T^\beta\wedge\eta_{\beta\mu\nu} +
T^\beta\wedge K^{\mu\nu}\,\eta_{\alpha\beta\mu\nu}\right).\label{TH1}
\end{equation}
Recalling that $T^\beta\wedge\eta_{\beta\mu\nu}=D\eta_{\mu\nu}$ and using
the fundamental identity (\ref{ID2}), we get
\begin{eqnarray}
{\frac 1 {2\kappa}}\,(e_{\alpha}\rfloor K^{\mu\nu})\,T^\beta\wedge
\eta_{\beta\mu\nu} &=& {\frac 1 {2\kappa}}\,(e_{\alpha}\rfloor K^{\mu\nu})\,
\vartheta_\mu\wedge K^{\rho\sigma}\wedge\eta_{\nu\rho\sigma}\nonumber\\
&=& (e_{\alpha}\rfloor T^{\beta})\wedge H^{(0)}_{\beta} + K_\alpha{}^\beta
\wedge H^{(0)}_\beta,\label{TH2}
\end{eqnarray}
where we used the Leibniz rule for the interior product, $(e_{\alpha}\rfloor 
K^{\mu\nu})\,\vartheta_\mu = e_{\alpha}\rfloor (K^{\mu\nu}\wedge\vartheta_\mu)
+ K_\alpha{}^\nu$, and identities (\ref{contor}) and (\ref{H0K}). Hence,
(\ref{TH1}) and (\ref{TH2}) yield
\begin{equation}
{\frac 1 {2\kappa}}\,T^\beta\wedge K^{\mu\nu}\,\eta_{\alpha\beta\mu\nu}\equiv
- K_\alpha{}^\beta\wedge H^{(0)}_\beta + T^\beta\wedge (e_{\alpha}\rfloor 
H^{(0)}_\beta) -(e_{\alpha}\rfloor T^{\beta})\wedge H^{(0)}_{\beta}.\label{TH3}
\end{equation}
On the other hand, 
\begin{equation}
T^\beta\,\eta_{\alpha\beta\mu\nu} = D\eta_{\alpha\mu\nu}= - K_\alpha{}^\beta
\wedge\eta_{\beta\mu\nu} - K_\mu{}^\beta\wedge\eta_{\alpha\beta\nu} -
K_\nu{}^\beta\wedge\eta_{\alpha\mu\beta}.
\end{equation}
Exterior product of this with ${\frac 1 {2\kappa}}\,K^{\mu\nu}$ yields,
taking into account (\ref{H0K}),
\begin{equation}
{\frac 1 {2\kappa}}\,T^\beta\wedge K^{\mu\nu}\,\eta_{\alpha\beta\mu\nu}\equiv
K_\alpha{}^\beta\wedge H^{(0)}_\beta - {\frac 1 \kappa}\,K^{\mu\nu}\wedge
K_\mu{}^\beta\wedge\eta_{\alpha\beta\nu}.\label{TH4}
\end{equation}
Comparing (\ref{TH3}) and (\ref{TH4}), we obtain:
\begin{eqnarray}
K_\alpha{}^\beta\wedge H^{(0)}_\beta &\equiv& {\frac 1 {2\kappa}}\,K^{\mu\nu}
\wedge K_\mu{}^\beta\wedge\eta_{\alpha\beta\nu} \nonumber\\
&& + {\frac 1 2}\,T^\beta\wedge (e_{\alpha}\rfloor H^{(0)}_\beta) - 
{\frac 1 2}\,(e_{\alpha}\rfloor T^{\beta})\wedge H^{(0)}_{\beta}.\label{KH}
\end{eqnarray}
Substituting (\ref{H0K}) and (\ref{KH}) into (\ref{DH0}), we finally get
\begin{eqnarray}
DH^{(0)}_{\alpha}&\equiv& {\frac 1 {2\kappa}}\,\left(\widetilde{D}K^{\mu\nu} -
K_\gamma{}^\mu\wedge K^{\nu\gamma}\right)\wedge\eta_{\alpha\mu\nu}\nonumber\\ 
&& - {\frac 1 2}\,T^\beta\wedge (e_{\alpha}\rfloor H^{(0)}_\beta) +
{\frac 1 2}\,(e_{\alpha}\rfloor T^{\beta})\wedge H^{(0)}_{\beta}.\label{DHf}
\end{eqnarray}

We are now in a position to compute the left hand side of the {\it first}
field equation. Collecting together (\ref{DHf}), (\ref{Ea0f}), and using
the decomposition of the curvature (\ref{RK}), one obtains
\begin{equation}
DH^{(0)}_{\alpha} - E^{(0)}_{\alpha}\equiv - \,{\frac 1 {2\kappa}}\,
\widetilde{R}^{\mu\nu}\wedge\eta_{\alpha\mu\nu}.\label{firstL}
\end{equation}

For the Lagrangian (\ref{V0lag}), the complete system of the {\it first} and 
{\it second} gravitational field equations thus reads:
\begin{eqnarray}
-\,{\frac 1 {2\kappa}}\,\widetilde{R}^{\mu\nu}\wedge\eta_{\alpha\mu\nu}&=&
\Sigma_\alpha,\label{first0}\\
0&=&\tau^\alpha{}_\beta.\label{second0}
\end{eqnarray}
This is the true Einstein's theory in which only the matter with the vanishing 
spin $\tau^\alpha{}_\beta=0$ is allowed, and the energy-momentum of matter 
$\Sigma_\alpha$ determines the {\it purely Riemannian geometry} via the
Einstein's equations (the components of the three-form on the left hand side of 
(\ref{first0}) comprise the standard Einstein tensor). 

The degenerate features of the model (\ref{V0lag}) are explained by the
existence of the {\it auxiliary} (or occasional) symmetry of the Lagrangian.
The symmetry in question arises from the transformation of the linear
connection {\it alone}: 
\begin{equation}
\delta_\varepsilon\Gamma_{\beta}{}^{\alpha}=\varepsilon_{\beta}{}^{\alpha},
\quad\quad \delta_\varepsilon\vartheta^\alpha=0. \label{traG}
\end{equation}
Here $\varepsilon_{\beta}{}^{\alpha}$ is an arbitrary tensor-valued one form,
antisymmetric in its indices $\varepsilon^{\alpha\beta}=-
\varepsilon^{\beta\alpha}$. Taking into account that $\delta_\varepsilon 
R_{\beta}{}^{\alpha}=D\varepsilon_{\beta}{}^{\alpha}$ and $\delta_\varepsilon
T^\alpha=\varepsilon_{\beta}{}^{\alpha}\wedge\vartheta^\beta$, we can
straightforwardly calculate the variation of the Lagrangian (\ref{V0lag}):
\begin{eqnarray}
\delta_\varepsilon V^{(0)}&=&{\frac 1 {2\kappa}}\,\delta_\varepsilon 
R_{\beta}{}^{\alpha}\wedge\eta^\beta{}_\alpha - \delta_\varepsilon T^\alpha
\wedge H^{(0)}_\alpha ={\frac 1 {2\kappa}}\left((D\varepsilon_{\beta}
{}^{\alpha})\wedge\eta^\beta{}_\alpha - \varepsilon_{\beta}{}^{\alpha}
\wedge D\eta^\beta{}_\alpha\right)\nonumber\\
&=&{\frac 1 {2\kappa}}\,d\left(\varepsilon_{\beta}{}^{\alpha}\wedge
\eta^\beta{}_\alpha\right).
\end{eqnarray}
The identities (\ref{H0K}) and (\ref{ID2}) were used above. Thus the 
{\it action} is unchanged $\delta_\varepsilon(\int V^{(0)})=0$, and we have 
demonstrated that the model (\ref{V0lag}) is {\it invariant} under the
transformation (\ref{traG}). This ``auxiliary'' symmetry has nothing to do 
with the linear or Poincar\'e gauge group underlying the gravity theory. The 
number of free parameters involved is $24$ ($\varepsilon^{\alpha\beta}$ is a 
one-form with skew symmetry). This is exactly equal to the number of 
components of the torsion (or contortion). In principle, it is possible to 
use the symmetry (\ref{traG}) and ``gauge away'' the torsion completely, 
transforming from the total connection to the {\it purely Riemannian} one:
\begin{equation}
\Gamma_{\beta}{}^{\alpha},\quad K_{\beta}{}^{\alpha}\neq 0\quad \longrightarrow
\quad \Gamma_{\beta}{}^{\alpha}=\widetilde{\Gamma}_{\beta}{}^{\alpha},\quad 
K_{\beta}{}^{\alpha}=0.
\end{equation}

The problem of the ``auxiliary" symmetry in the teleparallel and the 
Poincar\'e gauge gravity was discussed in 
\cite{Hecht,Kopc,Nester,Leclerc1,Leclerc2,Vadim2}.

\section{Double duality properties of the irreducible parts of the
Riemann-Cartan curvature}\label{doubleduality}

In this section we will demonstrate that the irreducible parts of the 
Riemann-Cartan curvature are all characterized by the {\it double-duality}
property
\begin{equation}
{}^{\star (I)}R_{\alpha\beta}=K_I\,{\frac 1 2}\,\eta_{\alpha\beta\mu\nu}\,
{}^{(I)}R^{\mu\nu},\qquad I = 1,\dots,6,\label{dd}
\end{equation}
where the double duality index $K_I$ is either $+1$ or $-1$. We will 
prove (\ref{dd}) and compute $K_I$ for each irreducible curvature part.

The terminology is explained as follows. In addition to the Hodge (``left")
duality operator, for the Lorentz algebra-valued objects $\psi_{\alpha\beta}
= - \psi_{\beta\alpha}$ we can define the ``right" duality operator by
\begin{equation}
\psi_{\alpha\beta}^\star := {\frac 1 2}\,\eta_{\alpha\beta\mu\nu}
\,\psi^{\mu\nu}.\label{rightd}
\end{equation}
Then, from (\ref{dd}) we find for the irreducible curvature parts
\begin{equation}
{}^{\star (I)}R_{\alpha\beta}^\star = - K_I\,
{}^{(I)}R^{\mu\nu},\qquad I = 1,\dots,6.\label{dd1}
\end{equation}
For $K_I = -1$ we have the {\it double dual} objects, whereas for $K_I = 1$
we find {\it anti-double dual} objects. 

\subsection{Mathematical preliminaries}

In our proof of (\ref{dd}) we will use the possibility of generating of the 
2-forms with the double duality properties from the different irreducible 
pieces of the vector-valued 1-forms. This fact can be formulated in terms 
of the following three lemmas.

{\bf Lemma A}: Let $A_\alpha$ be a vector-valued 1-form such that
\begin{equation}
A_\alpha\wedge\eta_\beta = - A_\beta\wedge\eta_\alpha.\label{condA}
\end{equation}
Then this 1-form satisfies the identity
\begin{equation}
{\frac 1 2}\eta^{\alpha\beta\mu\nu}\vartheta_\alpha\wedge A_\beta -
{}^\star\!\left(\vartheta^{[\mu}\wedge A^{\nu]}\right)\equiv 0.\label{idA}
\end{equation}

{\bf Lemma B}: Let $B_\alpha$ be a vector-valued 1-form such that
\begin{equation}
B_\alpha\wedge\eta_\beta = B_\beta\wedge\eta_\alpha,\quad\quad
B_\alpha\wedge\eta^\alpha=0.\label{condB}
\end{equation}
Then this 1-form satisfies the identity
\begin{equation}
{\frac 1 2}\eta^{\alpha\beta\mu\nu}\vartheta_\alpha\wedge B_\beta +
{}^\star\!\left(\vartheta^{[\mu}\wedge B^{\nu]}\right)\equiv 0.\label{idB}
\end{equation}

The proofs of these lemmas are given in the Appendix 2. The identities (\ref{idA}) 
and (\ref{idB}) mean that any 1-form $A_\alpha$ (resp., any 1-form $B_\alpha$) 
satisfying the conditions (\ref{condA}) (resp., (\ref{condB})) defines an {\it 
anti-double-dual} 2-form $\vartheta_{[\alpha}\wedge A_{\beta]}$ (resp., a {\it 
double-dual} 2-form $\vartheta_{[\alpha}\wedge B_{\beta]}$). 

Now we will show that the different irreducible pieces of an arbitrary vector-valued 
1-form generate the double dual 2-forms. At first, we recall that any 1-form 
$\rho_\alpha$ is decomposed into the sum
\begin{equation}\label{decrho}
\rho_\alpha = {}^{(1)}\rho_\alpha + {}^{(2)}\rho_\alpha + {}^{(3)}\rho_\alpha,
\end{equation}
where the irreducible parts are defined by
\begin{eqnarray}
{}^{(1)}\rho_\alpha &:=& \rho_\alpha - {\frac 1 4}\,\vartheta_\alpha
\,(e^\beta\rfloor\rho_\beta) - {\frac 1 2}\,e_\alpha\rfloor(\vartheta^\beta
\wedge \rho_\beta),\label{symrho}\\
{}^{(2)}\rho_\alpha &:=& {\frac 1 2}\,e_\alpha\rfloor(\vartheta^\beta
\wedge X_\beta),\label{antirho}\\
{}^{(3)}\rho_\alpha &:=& {\frac 1 4}\,\vartheta_\alpha\,(e^\beta\rfloor
\rho_\beta).\label{tracerho}
\end{eqnarray}
In tensor language, $\rho_\alpha$ describes the second rank tensor, and the
decomposition (\ref{decrho}) splits this tensor into the traceless symmetric
piece (\ref{symrho}), antisymmetric piece (\ref{antirho}), and the trace 
(\ref{tracerho}). 

{\bf Lemma C}: For an arbitrary 1-form $\rho_\alpha$, the 2-forms 
$\vartheta_{[\alpha}\wedge{}^{(2)}\rho_{\beta]}$ and $\vartheta_{[\alpha}
\wedge{}^{(3)}\rho_{\beta]}$ are anti-double-dual, whereas the 2-form
$\vartheta_{[\alpha}\wedge{}^{(1)}\rho_{\beta]}$ is double dual. 

Using the definition (\ref{symrho}), one immediately proves:
\begin{eqnarray}
{}^{(1)}\rho_\alpha\wedge\eta_\beta &=& \rho_\alpha\wedge\eta_\beta - 
{\frac 1 4}\vartheta_\alpha\wedge\eta_\beta\,(e^\gamma\rfloor\rho_\gamma)
- {\frac 1 2}(\rho_\alpha - \vartheta^\lambda 
e_\alpha\rfloor \rho_\lambda)\wedge\eta_\beta\nonumber\\
&=& {\frac 1 2}\,\rho_\alpha\wedge\eta_\beta - 
{\frac 1 4} g_{\alpha\beta}\,\eta\,(e^\gamma\rfloor\rho_\gamma)
+ {\frac 1 2}\eta\, e_\alpha\rfloor\rho_\beta\nonumber\\
&=&{\frac 1 2}(\rho_\alpha\wedge\eta_\beta + \rho_\beta\wedge\eta_\alpha)
- {\frac 1 4} g_{\alpha\beta}\,\eta\,(e^\gamma\rfloor\rho_\gamma).
\end{eqnarray}
Here we repeatedly used the basic identity
$\vartheta_\alpha\wedge\eta_\beta=g_{\alpha\beta}\,\eta.$
Hence, ${}^{(1)}\rho_\alpha$ satisfies the conditions (\ref{condB}):
\begin{equation}
{}^{(1)}\rho_\alpha\wedge\eta_\beta = {}^{(1)}\rho_\beta\wedge\eta_\alpha,
\qquad{}^{(1)}\rho_\alpha\wedge\eta^\alpha=0.\label{condPsi}
\end{equation}
Accordingly, by the Lemma B, the 2-form $\vartheta_{[\alpha}\wedge
{}^{(1)}\rho_{\beta]}$ is double dual.

As to the second irreducible part (\ref{antirho}), we find
\begin{equation}
{}^{(2)}\rho_\alpha\wedge\eta_\beta = {\frac 12}\left(\rho_\alpha\wedge
\eta_\beta - \eta\, e_\alpha\rfloor \rho_\beta\right) = {\frac 12}\left(
\rho_\alpha\wedge\eta_\beta - \rho_\beta\wedge\eta_\alpha\right),
\end{equation}
and thus the condition (\ref{condA}) of the Lemma A is explicitly fulfilled
for ${}^{(2)}\rho_\alpha$. 

The third irreducible part (\ref{tracerho}) is proportional to the 
coframe 1-form $\vartheta^\alpha$. The latter does not satisfy either 
(\ref{condA}) or (\ref{condB}). However, the coframe is directly
involved in the construction of Hodge duals, and, by definition,
\begin{equation}
{}^\star(\vartheta_\alpha\wedge\vartheta_\beta)=\eta_{\alpha\beta}=
{\frac 1 2}\eta_{\mu\nu\alpha\beta}\vartheta^\mu\wedge\vartheta^\nu.
\label{addvta1}
\end{equation}
Thus, the 2--form $\vartheta_\alpha\wedge\vartheta_\beta$ is {\it 
anti-double dual}, and the same is true for $\vartheta_{[\alpha}
\wedge{}^{(3)}\rho_{\beta]} = {\frac 14}(e^\gamma\rfloor\rho_\gamma)
\,\vartheta_\alpha\wedge\vartheta_\beta$. As a by-product of our
analysis, we note that 
\begin{equation}
{}^\star\eta_{\alpha\beta}={}^\star{}^\star(\vartheta_\alpha\wedge
\vartheta_\beta)=-\vartheta_\alpha\wedge\vartheta_\beta=
{\frac 1 2}\eta_{\mu\nu\alpha\beta}\eta^{\mu\nu},\label{addvta2}
\end{equation}
hence $\eta_{\alpha\beta}$ is {\it anti-double dual} too.

\subsection{Dual properties of the curvature}

Now we are in a position to prove the double duality properties (\ref{dd}).
As a first step, we notice that the definitions of the irreducible parts 
of the Riemann-Cartan curvature (\ref{curv2})-(\ref{curv1}) involve the
pair of the vector-valued 1-forms $W_\alpha$ and $X_\alpha$ defined in
(\ref{WX}). As a result, we can straightforwardly apply the lemmas A-C. 
Since $\Psi_\alpha = {}^{(1)}X_\alpha$ and $\Phi_\alpha = {}^{(1)}W_\alpha$,
see eqs. (\ref{Psia}) and (\ref{Phia}), we can immediately apply the Lemma
B to the 2-nd and the 4-th irreducible parts of the Riemann-Cartan curvature 
to demonstrate that these two pieces are both double dual. All the rest 
irreducible parts are anti-double dual. Indeed, for the 5-th part this is 
proved via the lemma A, using the fact that it involves ${}^{(2)}W_\alpha$, 
whereas for the 3-rd and the 6-th pieces this is obvious from (\ref{addvta1}) 
and (\ref{addvta2}). The proof for the Weyl type 1-st curvature piece is 
somewhat more nontrivial and will be given below. The complete list of the 
double duality properties for the Riemann-Cartan curvature reads:
\begin{eqnarray}
{}^{\star (1)}R_{\alpha\beta}&=&\, {\frac 1 2}\eta_{\alpha\beta\mu\nu}\,
{}^{(1)}R^{\mu\nu},\label{ddW1}\\
{}^{\star (2)}R_{\alpha\beta}&=& -{\frac 1 2}\eta_{\alpha\beta\mu\nu}\,
{}^{(2)}R^{\mu\nu},\label{ddW2}\\
{}^{\star (3)}R_{\alpha\beta}&=&\, {\frac 1 2}\eta_{\alpha\beta\mu\nu}\,
{}^{(3)}R^{\mu\nu},\label{ddW3}\\
{}^{\star (4)}R_{\alpha\beta}&=& -{\frac 1 2}\eta_{\alpha\beta\mu\nu}\,
{}^{(4)}R^{\mu\nu},\label{ddW4}\\
{}^{\star (5)}R_{\alpha\beta}&=&\, {\frac 1 2}\eta_{\alpha\beta\mu\nu}\,
{}^{(5)}R^{\mu\nu},\label{ddW5}\\
{}^{\star (6)}R_{\alpha\beta}&=&\, {\frac 1 2}\eta_{\alpha\beta\mu\nu}\,
{}^{(6)}R^{\mu\nu}.\label{ddW6}
\end{eqnarray}
As a comment to the proof of (\ref{ddW1}), we first notice that the
definition (\ref{curv1}) yields
\begin{eqnarray}
\eta^{\mu\nu}\wedge{}^{(1)}R_{\alpha\beta}&=& - {\frac 1 6}
\eta^{\mu\nu}{}_{\alpha\beta}\,\eta\, X - {\frac 1 3}
\delta^{[\mu}_{\alpha}\delta^{\nu]}_{\beta}\,\eta\, W +
2\,\eta^{[\mu}{}_{[\alpha}\wedge W^{\nu]}{}_{\beta]}\nonumber\\
&&-\delta^{[\mu}_{[\alpha}\left(\eta_{\beta]}\wedge W^{\nu]} +
\eta^{\nu]}\wedge W_{\beta]}\right).\label{etaW1}
\end{eqnarray}
{}From this we find $\eta^{\alpha\gamma}\wedge{}^{(1)}R_{\beta\gamma}=0$
(use the obvious identities $\eta_\alpha\wedge W^\beta = \eta_{\alpha
\gamma}\wedge W^{\beta\gamma}$ and $\eta_\alpha\wedge W^\alpha = 
- W\,\eta$). Since $(e_\mu\rfloor e_\nu\rfloor W_{\alpha\beta})\,\eta =
- \eta_{\mu\nu}\wedge W_{\alpha\beta}$ (use the Leibniz rule twice for
the interior product), the above two equations imply
\begin{equation}
e_\mu\rfloor e_\nu\rfloor{}^{(1)}R_{\alpha\beta}=
e_\alpha\rfloor e_\beta\rfloor{}^{(1)}R_{\mu\nu},\label{eeW1}
\end{equation}
and 
\begin{equation}
e_\mu\rfloor e_\nu\rfloor{}^{(1)}R^{\alpha\nu}=0.\label{eeW1c}
\end{equation}
Now, using the evident identity
\begin{equation}
{}^{(1)}R^{\alpha\beta}={\frac 1 2}\vartheta^\nu\wedge\vartheta^\mu\,
(e_\mu\rfloor e_\nu\rfloor{}^{(1)}R^{\alpha\beta}),
\end{equation}
one can straightforwardly compute
\begin{eqnarray}
{\frac 1 2}\eta_{\mu\nu\alpha\beta}\,{}^{\star (1)}R^{\alpha\beta}&=&
{\frac 1 8}\eta_{\mu\nu\alpha\beta}\eta^{\rho\sigma\delta\gamma}\,
\vartheta_\gamma\wedge\vartheta_\delta\, (e_\rho\rfloor e_\sigma\rfloor
{}^{(1)}R^{\alpha\beta})\nonumber\\
&=&{\frac 1 2}\vartheta_\gamma\wedge\vartheta_\delta\,
(e_\mu\rfloor e_\nu\rfloor{}^{(1)}R^{\gamma\delta})\nonumber\\
&=&{\frac 1 2}\vartheta_\gamma\wedge\vartheta_\delta\,
(e^\gamma\rfloor e^\delta\rfloor{}^{(1)}R_{\mu\nu})= - 
{}^{(1)}R_{\mu\nu},
\end{eqnarray}
where one have to expand the product $\eta_{\mu\nu\alpha\beta}\,
\eta^{\rho\sigma\delta\gamma}$ in terms of four Kronecker deltas,
then use (\ref{eeW1c}) to arrive at the second line, and subsequently use
(\ref{eeW1}) to get the final result (\ref{ddW1}).

A {\it simple corollary} of the double duality properties 
(\ref{ddW1})-(\ref{ddW6}) is the identity valid for any $I=1,...,6$:
\begin{equation}
(e_\alpha\rfloor{}^{(I)}R^{\mu\nu})\wedge{}^{*(I)}R_{\mu\nu}\equiv
{\frac 1 2}\,e_\alpha\rfloor\left({}^{(I)}R^{\mu\nu}\wedge{}^{*(I)}
W_{\mu\nu}\right),\label{zeroEaW}
\end{equation}
where for the sign coefficient $K_I$ see (\ref{ddW1})-(\ref{ddW6}), although
its value is in fact not important. Now we compute straightforwardly:
\begin{eqnarray}
(e_\alpha\rfloor{}^{(I)}R^{\mu\nu})\wedge{}^{*(I)}R_{\mu\nu}&=&
(e_\alpha\rfloor{}^{(I)}R^{\mu\nu})\wedge [K_I\,{\frac 1 2}\,
\eta_{\mu\nu\rho\sigma}\,{}^{(I)}R^{\rho\sigma}]\nonumber\\
&=&(e_\alpha\rfloor [K_I\,{\frac 1 2}\,\eta_{\mu\nu\rho\sigma}\,
{}^{(I)}R^{\mu\nu}])\wedge{}^{(I)}R^{\rho\sigma}\nonumber\\
&=&(e_\alpha\rfloor{}^{*(I)}R_{\rho\sigma})\wedge{}^{(I)}R^{\rho\sigma}
\nonumber\\ &=&e_\alpha\rfloor\left({}^{*(I)}R_{\rho\sigma}\wedge{}^{(I)}
W^{\rho\sigma}\right) - {}^{*(I)}R_{\rho\sigma}\wedge (e_\alpha\rfloor{}^{(I)}
W^{\rho\sigma}).
\end{eqnarray}
This concludes the proof of (\ref{zeroEaW}). 

Completely analogously, one can demonstrate that the following identities 
hold true among the elements of the two subsets of irreducible parts of 
curvature with the {\it same} duality coefficient $K_I$ (namely, within the subset 
$I=1,3,5,6$ with $K_I=1$ and within the subset $J=2,4$ with $K_J=-1$):
\begin{eqnarray}
(e_\alpha\rfloor{}^{(1)}R^{\mu\nu})\wedge{}^{*(3)}R_{\mu\nu}&=& 
(e_\alpha\rfloor{}^{(1)}R^{\mu\nu})\wedge{}^{*(6)}R_{\mu\nu}=0,\label{eaW136}\\
(e_\alpha\rfloor{}^{(1)}R^{\mu\nu})\wedge{}^{*(5)}R_{\mu\nu}&=&
(e_\alpha\rfloor{}^{(3)}R^{\mu\nu})\wedge{}^{*(5)}R_{\mu\nu}=0,\label{eaW135}\\
(e_\alpha\rfloor{}^{(3)}R^{\mu\nu})\wedge{}^{*(6)}R_{\mu\nu}&=&
(e_\alpha\rfloor{}^{(5)}R^{\mu\nu})\wedge{}^{*(6)}R_{\mu\nu}=0,\label{eaW356}\\
(e_\alpha\rfloor{}^{(2)}R^{\mu\nu})\wedge{}^{*(4)}R_{\mu\nu}&=&0.\label{eaW24} 
\end{eqnarray}
Among the elements belonging to the {\it different} subsets:
\begin{eqnarray}
(e_\alpha\rfloor{}^{(I)}R^{\mu\nu})\wedge{}^{*(J)}R_{\mu\nu}&=&
(e_\alpha\rfloor{}^{(J)}R^{\mu\nu})\wedge{}^{*(I)}R_{\mu\nu},\label{eaWsym}\\
{\rm for}\ I=1,3,5,6,\quad& {\rm and}&\ J=2,4.\nonumber
\end{eqnarray}
The identities (\ref{zeroEaW}) and (\ref{eaW136})-(\ref{eaWsym}) are 
extremely helpful in the computations for the gravitational energy-momentum
(\ref{Ea}) in the general quadratic models with the Lagrangians containing
the curvature square terms.

\section{Double duality solutions}

Let us consider now the general quadratic model (\ref{lagr}). The Double
Duality Ansatz (DDA) technique provides an effective method of finding
exact solutions of the field equations of the Poincar\'e gauge theory
(\ref{first}) and (\ref{second}). This method was developed in the numerous
papers \cite{Baekler1,Baekler2,Baekler3,Baekler4,Baekler5,Baekler6,Baekler7,Benn,Mielke1,Mielke2,Mielke3,Mielke4,Wallner,Vadim2}.

The general DDA represents the Lorentz gauge momentum in the form:
\begin{equation}
H^{\alpha}{}_{\beta}=\zeta\,{\frac 1 2}\,\eta^\alpha{}_{\beta\mu\nu}\,
R^{\mu\nu} - {\frac 1 {2\kappa}}\left(\xi\,\eta^\alpha{}_\beta + \chi\,
\vartheta^\alpha\wedge\vartheta_\beta\right),\label{DDA}
\end{equation}
where $\zeta, \xi, \chi$ are three constant parameters.

Let us consider in detail how the DDA works, separating the whole scheme 
into the simple steps, listed below in the following subsections.

\subsection{Second equation: solution for the translational momentum}

The exterior covariant derivative for (\ref{DDA}) is calculated 
straightforwardly:
\begin{eqnarray}
DH_{\alpha\beta}&=&\zeta\,{\frac 1 2}\,\eta_{\alpha\beta\mu\nu}\,
DR^{\mu\nu} - {\frac 1 {2\kappa}}\left(\xi\,D\eta_{\alpha\beta} + 
2\chi\,T_{[\alpha}\wedge\vartheta_{\beta]}\right)\nonumber\\
&=&\,-\,\xi\,\vartheta_{[\alpha}\wedge H^{(0)}_{\beta]} + {\frac \chi {\kappa}}
\,\vartheta_{[\alpha}\wedge T_{\beta]},\label{DHab1}
\end{eqnarray}
where we used the Bianchi identity $DR^{\mu\nu}\equiv 0$, and the 
fundamental identities (\ref{ID2}) and (\ref{H0K}). Substituting the
gravitational spin density (\ref{Eab}) into (\ref{second}), we obtain the 
{\it second} field equation in the form
\begin{equation}
-\,\xi\,\vartheta_{[\alpha}\wedge H^{(0)}_{\beta]} + {\frac \chi {\kappa}}\,
\vartheta_{[\alpha}\wedge T_{\beta]} + \vartheta_{[\alpha}\wedge H_{\beta]}
=\vartheta_{[\alpha}\wedge\mu_{\beta]}.\label{secondmu}
\end{equation}
Here we rewrote the matter source in terms of the {\it spin energy potential}
two-form $\mu_\alpha$ introduced in (\ref{taumu}) and (\ref{mutau}).
Equation (\ref{secondmu}) is formally solved with respect to the 
translational momentum:
\begin{equation}
H_\alpha = \xi\,H^{(0)}_\alpha - {\frac \chi {\kappa}}\,T_\alpha 
+ \mu_\alpha.\label{HaDDA}
\end{equation}
The analysis of this formal solution will be given in the subsequent 
Sec.~\ref{DDAtor}.

\subsection{First equation: reduction to the effective Einstein equation}

In terms of the gauge field momenta (\ref{Ha1}) and (\ref{Hab1}), the
Lagrangian (\ref{lagr}) reads:
\begin{equation}
V_{\rm Q}= {\frac {a_0} {4\kappa}}\,R^{\alpha\beta}\wedge\eta_{\alpha\beta}
- {\frac \lambda \kappa}\,\eta - {\frac 1 2}\,T^\alpha\wedge H_\alpha -
{\frac 1 2}\,R^{\alpha\beta}\wedge H_{\alpha\beta}.
\end{equation}
Inserting the DDA (\ref{DDA}) and the solution (\ref{HaDDA}), we find
\begin{eqnarray}
V_{\rm Q}&=& {\frac {a_0} {4\kappa}}\,R^{\alpha\beta}\wedge\eta_{\alpha\beta}
- {\frac \lambda \kappa}\,\eta - {\frac \zeta 4}\,\eta_{\alpha\beta\mu\nu}\,
R^{\alpha\beta}\wedge R^{\mu\nu} - {\frac 1 2}\,T^\alpha\wedge\mu_\alpha
\nonumber\\ &&+\,{\frac\xi{2\kappa}}\left({\frac 1 2}\,R^{\alpha\beta}\wedge
\eta_{\alpha\beta} - \kappa\,T^\alpha\wedge H^{(0)}_\alpha\right)\nonumber\\
&& +\,{\frac\chi{2\kappa}}\left({\frac 1 2}\,R^{\alpha\beta}\wedge\vartheta
_\alpha\wedge\vartheta_\beta + T^\alpha\wedge T_\alpha\right).\label{DDAlag}
\end{eqnarray}
In the similar way, we obtain
\begin{eqnarray}
(e_\alpha\rfloor T^\beta)\wedge H_\beta &=& \xi\,(e_\alpha\rfloor T^\beta)
\wedge H^{(0)}_\beta -{\frac\chi{2\kappa}}\,e_\alpha\rfloor (T^\beta\wedge 
T_\beta) + (e_\alpha\rfloor T^\beta)\wedge\mu_\beta,\label{DDATaHa}\\
(e_\alpha\rfloor R^{\mu\nu})\wedge H_{\mu\nu} &=& {\frac \zeta 4}\,
\eta_{\rho\sigma\mu\nu}\,e_\alpha\rfloor (R^{\rho\sigma}\wedge R^{\mu\nu})
\nonumber\\ && -{\frac 1{2\kappa}}\left[\xi\,(e_\alpha\rfloor R^{\mu\nu})\wedge
\eta_{\mu\nu} + \chi\,(e_\alpha\rfloor R^{\mu\nu})\wedge\vartheta_\mu\wedge
\vartheta_\nu\right].\label{DDARabHab}
\end{eqnarray}
Substituting (\ref{DDAlag}) and (\ref{DDATaHa})-(\ref{DDARabHab}) into
(\ref{Ea}), after some simple algebra we get the gravitational 
energy-momentum density:
\begin{eqnarray}
E_\alpha &=& {\frac 1{4\kappa}}\,e_\alpha\rfloor\left(a_0\,R^{\mu\nu}\wedge
\eta_{\mu\nu} - 4\lambda\,\eta - \xi\,R^{\mu\nu}\wedge\eta_{\mu\nu} -
\chi\,R^{\mu\nu}\wedge\vartheta_\mu\wedge\vartheta_\nu\right)\nonumber\\
&& +\,\xi\,E^{(0)}_\alpha + {\frac\chi\kappa}\,R_{\alpha\beta}\wedge
\vartheta^\beta + {\buildrel {(s)} \over E}_\alpha\nonumber\\
&=& -\,{\frac 1{4\kappa}}\,\eta_\alpha\left[4\lambda +(a_0 -\xi)\,W 
-\chi\,X \right] + \xi\,E^{(0)}_\alpha +{\frac\chi\kappa}\,R_{\alpha\beta}
\wedge\vartheta^\beta + {\buildrel {(s)} \over E}_\alpha.\label{EaDDA}
\end{eqnarray}
Here the {\it effective spin energy} three-form is introduced by
\begin{equation}
{\buildrel {(s)} \over E}_\alpha:={\frac 1 2}\left[(e_\alpha\rfloor T^\beta)
\wedge\mu_\beta - T^\beta\wedge (e_\alpha\rfloor\mu_\beta)\right].\label{ES}
\end{equation}
We used (\ref{Ea0f}) at the intermediate stage, and inserted the
contractions $R^{\mu\nu}\wedge\eta_{\mu\nu}={}^{(6)}R^{\mu\nu}\wedge
\eta_{\mu\nu}=-W\,\eta$ and $R^{\mu\nu}\wedge\vartheta_\mu\wedge\vartheta_\nu
={}^{(3)}R^{\mu\nu}\wedge\vartheta_\mu\wedge\vartheta_\nu=-X\,\eta$. The scalar
functions $W$ and $X$ are the Riemann-Cartan curvature scalar and pseudoscalar,
respectively. The covariant exterior differential of (\ref{HaDDA}) is 
\begin{equation}
DH_\alpha = \xi\,DH^{(0)}_\alpha + {\frac\chi\kappa}\,R_{\alpha\beta}\wedge
\vartheta^\beta,\label{DHaDDA}
\end{equation}
where the first Bianchi identity $DT_\alpha\equiv R_{\alpha\beta}\wedge
\vartheta^\beta$ was inserted.

Finally, making use of the identity (\ref{firstL}), we find the {\it first} 
gauge field equation (\ref{first}) in the form of the {\it effective 
Einstein} equation:
\begin{equation}
-\,{\frac \xi {2\kappa}}\,\widetilde{R}^{\mu\nu}\wedge\eta_{\alpha\mu\nu} + 
{\frac {\Lambda_{\rm eff}}\kappa}\,\eta_\alpha\,=
\Sigma_\alpha^{\rm eff}.\label{einDDA}
\end{equation}
Here we denote the effective energy-momentum of matter and the effective
cosmological term:
\begin{eqnarray}
\Sigma_\alpha^{\rm eff}&:=&\Sigma_\alpha + {\buildrel {(s)} \over E}_\alpha,
\label{Sigeff}\\
\Lambda_{\rm eff}&:=&\lambda +{\frac 1 4}\,[(a_0-\xi)\,W -\chi\,X].\label{Leff}
\end{eqnarray}
In general, the Riemann-Cartan curvature scalar $W$ and pseudoscalar $X$
are not constant, but the algebraic conditions on the curvature force them
to be constant, see below. The specific combination (\ref{Leff}) {\it must} 
be constant {\it in vacuum} ($\Sigma_\alpha=\tau^\alpha{}_\beta=0$) since 
taking covariant Riemannian exterior derivative of (\ref{einDDA}), 
we find $\widetilde{D}(\Lambda_{\rm eff}\,\eta_\alpha)=d\Lambda_{\rm eff}
\wedge\eta_\alpha=0$, hence $\Lambda_{\rm eff}$=const.

\subsection{Algebraic conditions on torsion}\label{DDAtor}

The formal solution (\ref{HaDDA}) for the translational gauge momentum
represents an algebraic system on the components of the torsion. We can write
it down explicitly substituting (\ref{Ha1}), (\ref{Ha0}) into (\ref{HaDDA}), 
and using the irreducible decomposition of the dual torsion,
\begin{equation}
{}^{(1)}({}^*T^\alpha)={}^*({}^{(1)}T^\alpha),\quad
{}^{(2)}({}^*T^\alpha)={}^*({}^{(3)}T^\alpha),\quad
{}^{(3)}({}^*T^\alpha)={}^*({}^{(2)}T^\alpha).
\end{equation}
The resulting system of the irreducible parts of the equation (\ref{HaDDA}) reads
\begin{eqnarray}
(a_1 + \xi)\,{}^{(1)}T^\alpha - \chi\,{}^*({}^{(1)}T^\alpha)
&=& -\,{}^{(1)}({}^*\mu^\alpha),\label{DDAT1}\\
(a_2 -2\xi)\,{}^{(2)}T^\alpha - \chi\,{}^*({}^{(3)}T^\alpha)
&=& -\,{}^{(2)}({}^*\mu^\alpha),\label{DDAT2}\\
\left(a_3 -{\frac\xi 2}\right)\,{}^{(3)}T^\alpha -\chi\,{}^*({}^{(2)}T^\alpha)
&=& -\,{}^{(3)}({}^*\mu^\alpha).\label{DDAT3}
\end{eqnarray}
Here the irreducible parts of the spin potential two-form ${}^{(I)}({}^*\mu
^\alpha)$ are defined in the same way as for the torsion. 

In generic {\it non-vacuum} case, when all irreducible parts of spin are
nontrivial, the choice of the constants on the left hand sides must allow
for the unique torsion solution. In particular, (\ref{DDAT1}) yields the
traceless irreducible part of torsion:
\begin{equation}
{}^{(1)}T^\alpha={\frac 1 {(a_1 + \xi)^2 + \chi^2}}\left[\chi\,{}^{(1)}
\mu^\alpha - (a_1 + \xi)\,{}^{*(1)}\mu^\alpha\right].\label{T1solDDA}
\end{equation}
Certainly, the denominator should be nonzero. Recall that until now we  
have not fixed the constant parameters $\xi$ and $\chi$ which enter the 
original DDA representation (\ref{DDA}). At this stage, it is enough just
to assume that $\chi\neq0$ and this guarantees the nonzero denominator
in (\ref{T1solDDA}) for any choice of $\xi$ and any value of the coupling
constant $a_1$. Analogously, we obtain from (\ref{DDAT2})-(\ref{DDAT3})
the trace and axial trace irreducible parts of torsion:
\begin{eqnarray}
{}^{(2)}T^\alpha &=& {\frac 1 {(a_2 - 2\xi)(a_3 -{\frac\xi 2}) + \chi^2}}
\left[\chi\,{}^{(2)}\mu^\alpha - \left(a_3 - {\frac\xi 2}\right)\,
{}^{*(3)}\mu^\alpha\right],\label{T2solDDA}\\
{}^{(3)}T^\alpha &=& {\frac 1 {(a_2 - 2\xi)(a_3 -{\frac\xi 2}) + \chi^2}}
\left[\chi\,{}^{(3)}\mu^\alpha - (a_2 - 2\xi)\,{}^{*(2)}\mu^\alpha
\right].\label{T3solDDA}
\end{eqnarray}
Here again the denominator must be nonzero which can always be guaranteed
by the proper choice of the constant parameters $\xi$ and $\chi$. 

A special word is necessary about the {\it vacuum DDA} solutions. When 
$\Sigma_\alpha=0, \tau_{\alpha\beta}=0$, hence $\mu_\alpha=0$, the 
generic solution of the system (\ref{DDAT1})-(\ref{DDAT3}) is $T^\alpha=0$,
which is clearly described by the formulas (\ref{T1solDDA})-(\ref{T3solDDA}).
Hence, the geometry becomes purely Riemannian with the metric determined 
from the vacuum Einstein equation (\ref{einDDA}).

The {\it nontrivial vacuum torsion} is only possible when 
\begin{equation}
(a_1 + \xi)^2 + \chi^2=0,\label{deg1}
\end{equation}
or/and
\begin{equation}
(a_2 - 2\xi)(a_3 -{\frac\xi 2}) + \chi^2=0.\label{deg23}
\end{equation}
With such a special choice of coefficients many DDA solutions were obtained
in the literature. However, it was immediately noticed that most of these 
solutions involve {\it free} functions which means that the torsion 
configurations are {\it not} determined unambiguously by the physical sources. 
This observation had stirred a confusion among the gravitational community
\cite{Mielke2,Hecht,Kopc,Nester}: indeed, how can 
(at least part of) the torsion be nondynamical and hence arbitrary when one 
``apparently'' can measure the torsion with the help of the particles with 
spin? So (citing the title of the paper \cite{Hecht}), ``can Poincar\'e 
gauge theory be saved?''  

Quite fortunately, the theory cures itself due to the self-consistency of 
its general scheme. Indeed, one simply has to recall what is a measurement
in a physical theory. For example, if we want to measure the torsion, what 
do we need for this? Clearly, we need a ``measuring device'' which feels 
the torsion. In physical terms, ``to feel'' means ``to interact with''. 
Thus we are again returning from the vacuum case to the theory with sources. 
Note that, evidently, one cannot choose one set of the coupling constants 
$a_I, b_J$ for vacuum and a different set for nontrivial sources. One must 
keep the coupling constants $a_I, b_J$ fixed in both cases, working within
one and the same particular model. 

To be specific, let us consider the vacuum DDA solutions which allow for
a nontrivial tracefree torsion ${}^{(1)}T^\alpha$. In vacuum, this is only
possible when (\ref{deg1}) is fulfilled. Hence one must put $a_1 + \xi=\chi
=0$. But in turn, such a choice (see (\ref{DDAT1})) means that the 
tracefree part of spin ${}^{(1)}\mu_\alpha=0$ always! To put it in a
different way, no ``measuring device'' which feels ${}^{(1)}T^\alpha$
is allowed in this model. Hence, ${}^{(1)}T^\alpha$ is {\it unobservable},
and there is no reason to worry about free functions which may occur in the
solutions: one cannot measure these configurations anyway.

The Dirac particles with spin ${\frac 1 2}$ represent the matter source which 
appears to be the most suitable for the measurement of torsion. As we know, the
Dirac spin is totally antisymmetric, which in terms of the spin energy
potential means that only the second irreducible part ${}^{(2)}\mu_\alpha$
is nontrivial and ${}^{(1)}\mu_\alpha={}^{(3)}\mu_\alpha=0$. [There is no
misprint: axial torsion is described by ${}^{(3)}T^\alpha$ whereas axial
spin is ${}^{(2)}\mu_\alpha$]. One can demonstrate that the vacuum DDA 
solutions with the vanishing axial torsion (which is consistent with the
equation (\ref{T3solDDA})) involve free functions in the trace and trace-free
torsion parts. In order to allow for such solutions one should restrict the
choice of the constants to (\ref{deg1})-(\ref{deg23}). However such a choice
then demands ${}^{(1)}\mu_\alpha={}^{(3)}\mu_\alpha=0$, and hence again the
nondynamical torsion parts are truly unobservable: spin ${\frac 1 2}$ particles
cannot detect them.

\subsection{Algebraic conditions on Riemann-Cartan curvature}\label{DDAcur}

Similarly, the double duality ansatz itself (\ref{DDA}) represents an
algebraic system on the Riemann-Cartan curvature. Here we analyze this 
system. At first, we substitute the explicit Lorentz gauge momentum (\ref{Hab1})
into (\ref{DDA}), and use the double duality properties 
(\ref{ddW1})-(\ref{ddW6}) for the Riemann-Cartan curvature. Then the 
irreducible parts of (\ref{DDA}) read as follows:
\begin{eqnarray}
(b_1 - \zeta)\,{}^{(1)}R_{\alpha\beta} &=& 0,\label{DDAw1}\\
(b_2 + \zeta)\,{}^{(2)}R_{\alpha\beta} &=& 0,\label{DDAw2}\\
(b_4 + \zeta)\,{}^{(4)}R_{\alpha\beta} &=& 0,\label{DDAw4}\\
(b_5 - \zeta)\,{}^{(5)}R_{\alpha\beta} &=& 0,\label{DDAw5}
\end{eqnarray}
for the traceless (1st, 2nd, 4th and 5th) parts of the curvature, and we find 
it convenient to write the trace and pseudotrace parts (6th and 3rd) 
separately,
\begin{eqnarray}
{\frac {(b_3 - \zeta)}6}\,X - {\frac\chi\kappa} &=& 0,\label{DDA-X}\\
{\frac {(b_6 - \zeta)}6}\,W + {\frac {(a_0-\xi)} \kappa} &=& 0.\label{DDA-W}
\end{eqnarray}

Since we still have one free parameter of DDA, namely $\zeta$, one can 
choose it in such a way that one of the coefficients in 
(\ref{DDAw1})-(\ref{DDA-W}) vanishes. Usual choice is $\zeta=-b_4$ which
eliminates the contribution of the fourth irreducible curvature part. In the
generic case, when no other coefficients vanish, one have to use the
remaining equations as the constraints on the components of the nontrivial
torsion. 

\section{A torsion kink}\label{kinky}

The DDA technique works also for non-vacuum solutions. As a particular
example \cite{Baekler7}, let us consider the gauge gravity coupled to the Higgs-type 
massless scalar field $\varphi$. The latter is described by the Lagrangian
\begin{equation}
L_{\rm mat}={\frac 1 2}\,d\varphi\wedge{}^*d\varphi.\label{Lhiggs}
\end{equation}
The total Lagrangian $V_{\rm Q} + L_{\rm mat}$ yields the non-vacuum
field equations (\ref{first}) and (\ref{second}) with the sources:
\begin{eqnarray}
\Sigma_\alpha &=& {\frac {\delta L_{\rm mat}} {\delta\vartheta^\alpha}}=
-\,{\frac 1 2}\left[(e_\alpha\rfloor d\varphi){}^*d\varphi +
d\varphi\wedge(e_\alpha\rfloor{}^*d\varphi)\right],\label{momhiggs}\\
\tau^\alpha{}_\beta &=& {\frac {\delta L_{\rm mat}} 
{\delta\Gamma_\alpha{}^\beta}}=0.\label{spinhiggs}
\end{eqnarray}
The vanishing spin (\ref{spinhiggs}) evidently leads to ${\buildrel {(s)} 
\over E}_\alpha=0$ in the effective Einstein equation. 

Besides, the matter (Klein-Gordon) field equation arises from the variation
of (\ref{Lhiggs}) with respect to the scalar field:
\begin{equation}
d\,{}^*d\varphi=0.\label{klein}
\end{equation}

Let us look for the spherically symmetric solution within the DDA approach.
We introduce the standard coordinate system $(t,r,\theta,\phi)$, and
assume the spherically symmetric ansatz for the coframe
\begin{equation}
\vartheta^{\hat{0}}=e^{\mu(r)}\,dt,\quad \vartheta^{\hat{1}}=e^{\nu(r)}\,dr,
\quad \vartheta^{\hat{2}}=r\,d\theta,\quad\vartheta^{\hat{3}}=r\,\sin\theta\,
d\phi.\label{metkink}
\end{equation}
The functions $\mu=\mu(r)$ and $\nu=\nu(r)$ depend only on the radial
coordinate $r$, as well as the scalar field $\varphi=\varphi(r)$. Substituting
this into (\ref{momhiggs}) and subsequently into the effective Einstein
equation (\ref{einDDA}), we find the following system:
\begin{eqnarray}
{\frac \kappa 2}\,(\varphi')^2 - 2{\frac {\nu'} r} + {\frac {1 - e^{2\nu}}
{r^2}} + \Lambda_{\rm eff}\,e^{2\nu} &=& 0,\label{einkink0}\\
{\frac \kappa 2}\,(\varphi')^2 - 2{\frac {\mu'} r} - {\frac {1 - e^{2\nu}}
{r^2}} - \Lambda_{\rm eff}\,e^{2\nu} &=& 0,\label{einkink1}\\
{\frac \kappa 2}\,(\varphi')^2 + \mu'' + (\mu' - \nu')\left(\mu' + {\frac 1 r}
\right) + \Lambda_{\rm eff}\,e^{2\nu} &=& 0.\label{einkink23}
\end{eqnarray}
Primes denote differentiation with respect to $r$. The Klein-Gordon equation
(\ref{klein}) yields 
\begin{equation}
\varphi'' + \left(\mu' - \nu' + {\frac 2 r}\right)\varphi'=0.\label{KGkink}
\end{equation}
{}From (\ref{einkink0}) and (\ref{einkink1}) we obtain
\begin{equation}
{\frac \kappa 2}\,(\varphi')^2 ={\frac {\mu' + \nu'} r},\label{munu1}
\end{equation}
and the scalar field equation (\ref{KGkink}) gives
\begin{equation}
\varphi'=\,C\,{\frac {e^{\nu-\mu}} {r^2}},\label{munu2}
\end{equation}
where $C$ is an integration constant.

We will not analyze the general solutions of the couple Einstein-Klein-Gordon
system. Let us confine ourselves to the particular case with
\begin{equation}
\mu=0.
\end{equation}
Then (\ref{einkink23}) and (\ref{munu1}) demand $\Lambda_{\rm eff}=0$, 
whereas (\ref{einkink0})-(\ref{einkink1}) yield the solution
\begin{equation}
e^{2\nu}=\left({1 + {\frac {C^2}{r^2}}}\right)^{-1}.\label{nusol}
\end{equation}
Here $C$ is the same integration constant as in (\ref{munu2}). Finally, 
equation (\ref{munu2}) is solved for the scalar field:
\begin{equation}
\varphi=\sqrt{\frac 2 \kappa}\,\left(\pm{\rm arcsinh}\left[{\frac C r}\right] 
+ C_1 \right).\label{scalsol}
\end{equation}

Spherically symmetric static torsion configurations are described by the 
general ansatz:
\begin{eqnarray}
T^{\hat{0}}&=&f\,\vartheta^{\hat{0}}\wedge\vartheta^{\hat{1}},\label{T0kink}\\
T^{\hat{1}}&=&h\,\vartheta^{\hat{0}}\wedge\vartheta^{\hat{1}},\label{T1kink}\\
T^{\hat{2}}&=&{\frac 1 r}\left(k\,\vartheta^{\hat{0}} 
+ g\,\vartheta^{\hat{1}}\right)\wedge\vartheta^{\hat{2}},\label{T2kink}\\
T^{\hat{3}}&=&{\frac 1 r}\left(k\,\vartheta^{\hat{0}} 
+ g\,\vartheta^{\hat{1}}\right)\wedge\vartheta^{\hat{3}},\label{T3kink}
\end{eqnarray}
where the functions $f=f(r), h=h(r), k=k(r), g=g(r)$ depend only on the
radial coordinate. Given the coframe (\ref{metkink}) and the torsion
(\ref{T0kink})-(\ref{T3kink}), it is straightforward to calculate the 
Riemann-Cartan curvature. Let us introduce, for convenience, the functions
\begin{equation}
F:=f + e^{-\nu}\,\mu',\quad\quad G:=g - e^{-\nu}.
\end{equation}
The direct calculation gives the 1st (Weyl) irreducible curvature part:
\begin{eqnarray}
{}^{(1)}R_{\hat{0}\hat{1}}&=& {\frac 1 3}\,U\,
\vartheta^{\hat{0}}\wedge\vartheta^{\hat{1}},\quad\quad
{}^{(1)}R_{\hat{2}\hat{3}}=-\,{\frac 1 3}\,U\,
\vartheta^{\hat{2}}\wedge\vartheta^{\hat{3}},\\
{}^{(1)}R_{\hat{0}\hat{2}}&=& -\,{\frac 1 6}\,U\,
\vartheta^{\hat{0}}\wedge\vartheta^{\hat{2}},\quad
{}^{(1)}R_{\hat{3}\hat{1}}=\,{\frac 1 6}\,U\,
\vartheta^{\hat{3}}\wedge\vartheta^{\hat{1}},\\
{}^{(1)}R_{\hat{0}\hat{3}}&=& -\,{\frac 1 6}\,U\,
\vartheta^{\hat{0}}\wedge\vartheta^{\hat{3}},\quad
{}^{(1)}R_{\hat{1}\hat{2}}=\,{\frac 1 6}\,U\,
\vartheta^{\hat{1}}\wedge\vartheta^{\hat{2}},
\end{eqnarray}
where
\begin{equation}
U=e^{-\nu}\left(F' + F\mu' + {\frac {G'} r}\right) + {\frac 1 {r^2}}\left(
(FG + hk)\,r + G^2 - k^2 -1\right).\label{Ukink}
\end{equation}
The 2nd and the 3rd parts are trivial,
\begin{equation}
{}^{(2)}R_{\alpha\beta}={}^{(3)}R_{\alpha\beta}=0\,,\label{zeroW23}
\end{equation}
whereas the 5th part reads
\begin{equation}
{}^{(5)}R_{\alpha\beta}=-\,\vartheta_{[\alpha}\wedge
e_{\beta]}\rfloor\Phi\,,
\end{equation}
where the 2-form 
\begin{equation}
\Phi:={\frac 1 r}\left(e^{-\nu}\,k' + Fk + Gh\right)\vartheta^{\hat{0}}
\wedge\vartheta^{\hat{1}}.\label{phikink}
\end{equation}
The 6th irreducible part:
\begin{equation}
{}^{(6)}R_{\alpha\beta}\,= -\,{\frac 1 {12}}\,W\,
\vartheta_{\alpha}\wedge\vartheta_{\beta},
\end{equation}
where the curvature scalar is
\begin{equation}
W=2U -\,{\frac 6 r}\left(e^{-\nu}\,G' + FG + kh\right).\label{Wkink}
\end{equation}
For completeness, let us write down the 4th irreducible part of curvature
(\ref{curv4}) which describes the traceless symmetric Ricci tensor
(\ref{Ric}). The corresponding one-form $\Phi_{\alpha}$ has the 
following components
\begin{eqnarray}
\Phi_{\hat{0}}&=& (A + P)\,\vartheta^{\hat{0}} +
B\,\vartheta^{\hat{1}},\\
\Phi_{\hat{1}}&=& (A - P)\,\vartheta^{\hat{1}} +
B\,\vartheta^{\hat{0}},\\
\Phi_{\hat{2}}&=& P\,\vartheta^{\hat{2}},\qquad
\Phi_{\hat{3}}= P\,\vartheta^{\hat{3}},
\end{eqnarray}
where we denoted
\begin{eqnarray}
A &:=& {\frac 1 r}\,(e^{-\nu}G' - FG + kh),\\
B &:=& {\frac 1 r}\,(e^{-\nu}k' - Fk + Gh),\\
P &:=& {\frac 1 2}\,\left[e^{-\nu}(F' + F\mu') + {\frac 1 {r^2}}(-G^2 + k^2 +1)
\right].
\end{eqnarray}

Now we are in a position to solve the algebraic curvature equations 
(\ref{DDAw1})-(\ref{DDA-W}). We choose the DDA parameter $\zeta=-b_4$, 
which makes (\ref{DDAw4}) automatically satisfied. The remaining equations
yield vanishing of the 1st and 5th irreducible parts (recall that 2nd and 
3rd are already zero, (\ref{zeroW23}), hence $\chi=0$), and the constancy of 
the curvature scalar. Denote the constant
\begin{equation}
A_0 := {\frac {a_0 - \xi} {\kappa(b_4 + b_6)}}.
\end{equation}
{}From (\ref{DDA-W}) we have $W=-6A_0$, and (\ref{Ukink}), (\ref{phikink}),
(\ref{Wkink}) yield the final differential system:
\begin{eqnarray}
e^{-\nu}\,(F' + F\mu') + {\frac 1 {r^2}}\,(G^2 - k^2 -1) &=& -A_0,\\
e^{-\nu}\,k' + Fk + Gh &=& 0,\\
e^{-\nu}\,G' + FG + kh &=& A_0\,r.
\end{eqnarray}
We will not analyze the complete solution of this system. Instead, consider
a particular solution:
\begin{equation}
f=g=0,\quad\quad h=k',\quad\quad k=\pm\sqrt{A_0\,r^2 + {\frac {C^2}{r^2}}}.
\end{equation}
We can verify for the effective (zero) cosmological constant that
\begin{equation}
\Lambda_{\rm eff}=\lambda - {\frac 3 2}
{\frac {(a_0-\xi)^2} {\kappa(b_4+b_6)}}=0.
\end{equation}

\section{Torsion-free solutions}\label{vacuum}

The numerous classical exact and approximate solutions (including the demonstration 
of the generalized Birkhoff theorem for the spherical symmetry) were derived in
\cite{Garecki,Glad,Mccrea1,Mccrea2,Mink,Rama,Rauch,Vadim1,Vadim2,Vadim3}, 
to mention but a few papers. The detailed overview can be found, for example, in
\cite{OPZ}. 

Let us consider the vacuum solutions with {\it vanishing torsion} in the
general quadratic models (\ref{lagr}). For $T^\alpha=0$, from (\ref{Ha1})
and (\ref{Eab}) we immediately find (to recall, the tilde denotes the torsion-free
Riemannian objects)
\begin{equation}
\widetilde{H}_\alpha=0,\quad\quad \widetilde{E}^\alpha{}_\beta=0.
\end{equation}
The only nontrivial irreducible parts of the curvature are the Weyl form
${}^{(1)}\widetilde{W}^{\mu\nu}$, the traceless Ricci form ${}^{(4)}
\widetilde{W}^{\mu\nu}$ and the curvature scalar ${}^{(6)}\widetilde{W}
^{\mu\nu}=-{\frac 1 {12}}\widetilde{W}\vartheta^\mu\wedge\vartheta^\nu$. The 
Lorentz gauge momentum (\ref{Hab1}) reads
\begin{equation}
\widetilde{H}_{\alpha\beta}=-\,{\frac {a_0} {2\kappa}} + b_1\,{}^{*(1)}
\widetilde{W}_{\alpha\beta} + b_4\,{}^{*(4)}\widetilde{W}_{\alpha\beta} + 
b_6\,{}^{*(6)}\widetilde{W}_{\alpha\beta},
\end{equation}
and hence the second vacuum equation (\ref{second}) reduces to
\begin{equation}
b_1\,\widetilde{D}({}^{*(1)}\widetilde{W}_{\alpha\beta}) + b_4\,\widetilde{D}
({}^{*(4)}\widetilde{W}_{\alpha\beta}) + b_6\,\widetilde{D}({}^{*(6)}
\widetilde{W}_{\alpha\beta})=0.\label{secondR1}
\end{equation}
This can be simplified with the help of the Bianchi identity 
\begin{equation}
\widetilde{D}\widetilde{R}^{\mu\nu}\equiv\widetilde{D}\,{}^{(1)}\widetilde{W}
^{\mu\nu} + \widetilde{D}\,{}^{(4)}\widetilde{W}^{\mu\nu} + \widetilde{D}\,
{}^{(6)}\widetilde{W}^{\mu\nu}\equiv 0.
\end{equation} 
Contracting the last identity with ${\frac 1 2}\eta_{\alpha\beta\mu\nu}$ and
using the double duality properties (\ref{ddW1})-(\ref{ddW6}), we obtain
\begin{equation}
\widetilde{D}({}^{*(1)}\widetilde{W}_{\alpha\beta}) - \widetilde{D}
({}^{*(4)}\widetilde{W}_{\alpha\beta}) + \widetilde{D}({}^{*(6)}
\widetilde{W}_{\alpha\beta})\equiv 0.
\end{equation}
We can eliminate the derivative of the Weyl form in (\ref{secondR1}), and
write the {\it second} field equation as
\begin{equation}
(b_1 + b_4)\,\widetilde{D}({}^{*(4)}\widetilde{W}_{\alpha\beta}) + {\frac 1 
{12}}(b_1 - b_6)\,d\widetilde{W}\wedge\eta_{\alpha\beta}=0,\label{secondR2}
\end{equation}
where we substituted the 6th irreducible curvature explicitly in terms of
the curvature scalar $\widetilde{W}$.

The gravitational energy (\ref{Ea}) is calculated straightforwardly with
the help of the identities (\ref{zeroEaW}) and (\ref{eaW136})-(\ref{eaWsym}),
and the {\it first} vacuum field equation is written in the form:
\begin{eqnarray}
-\widetilde{E}_\alpha &=& -{\frac {a_0}{2\kappa}}\,\widetilde{R}^{\mu\nu}
\wedge\eta_{\alpha\mu\nu} + {\frac \lambda\kappa}\,\eta_\alpha \nonumber\\
&& + (b_1 + b_4)\,\widetilde{\Phi}^\beta\wedge{}^{*(1)}\widetilde{W}
_{\alpha\beta} - {\frac 1 6}(b_4 + b_6)\,\widetilde{W}\,
{}^*\widetilde{\Phi}_\alpha\,=0.\label{firstR1}
\end{eqnarray}
Here we used the explicit representation of the 4th irreducible curvature
part ${}^{(4)}\widetilde{W}_{\alpha\beta}=-\vartheta_{[\alpha}\wedge
\Phi_{\beta]}$ in terms of the one-form (\ref{Phia}). Note that 
$\widetilde{R}^{\mu\nu}\wedge\eta_{\alpha\mu\nu}\equiv 2{}^*\widetilde{\Phi}
_\alpha - {\frac 1 2}\,\widetilde{W}\,\eta_\alpha$. 

One can somewhat simplify the resulting field equations. Transvecting 
(\ref{firstR1}) with $\vartheta^\alpha$, we find that the curvature
scalar is constant
\begin{equation}
\widetilde{W} = -\,{\frac {4\lambda} {a_0}}.\label{scalR}
\end{equation}
(For the purely quadratic models with $a_0=0$ the cosmological term should
also be zero $\lambda=0$). Hence the last term in (\ref{secondR2}) vanishes.
Substituting (\ref{scalR}) back into (\ref{secondR2})-(\ref{firstR1}) we
obtain the final system of algebraic-differential equations:
\begin{eqnarray}
(b_1 + b_4)\,\widetilde{\Phi}^\beta\wedge{}^{*(1)}\widetilde{W}_{\alpha\beta}
&=& \left({\frac {a_0} \kappa} - {\frac {2\lambda} {3a_0}}(b_4 + b_6)\right)
{}^*\widetilde{\Phi}_\alpha, \label{firstR}\\
(b_1 + b_4)\,\widetilde{D}(e_{[\alpha}\rfloor{}^*\widetilde{\Phi}_{\beta]})
&=& 0.\label{secondR}
\end{eqnarray}
All the solutions of the vacuum Einstein equations with a cosmological term
\begin{equation}
\widetilde{\Phi}_\alpha=0,\label{ein}
\end{equation}
see (\ref{Ric}), are evidently also the torsion-free solutions of the general 
quadratic Poincar\'e gauge models. One can prove \cite{Debney,Fair1,Fair2} that 
the Einstein spaces (\ref{ein}) are {\it the only} torsion-free vacuum solutions 
of (\ref{firstR})-(\ref{secondR}) except for the three very specific degenerate 
choices of the coupling constants \cite{OPZ}:
\begin{equation}
b_6 - {\frac {3a_0^2}{2\lambda\kappa}} = \left\{\begin{array}{c} 
b_1, \\ - b_4,\\ -2b_1 - 3b_4.\end{array}\right.\label{except}  
\end{equation}

\section{Conclusion}\label{conclude}

In this paper, an overview of the selected aspects of the Poincar\'e gauge gravity
is given. The Lagrange-Noether approach is formulated in a general way and the
conservation laws and the field equations are derived. As a particular 
application, we analyze the family of quadratic (in the curvature and the 
torsion) models. The new results obtained include the discussion of the special 
case of the spinless matter and the demonstration that Einstein's theory arises 
as a degenerate model in the class of the quadratic Poincar\'e theories. Finally,
we outlined the main features of the so-called double duality method for 
constructing of the exact solutions of the quadratic Poincar\'e gauge theories. 

\bigskip
{\bf Acknowledgments}. This paper is the outcome of the joint work with Friedrich Hehl
at the University of Cologne. It was supported, at different stages, by the funds
from the Alexander-von-Humboldt Foundation (Bonn), the Deutsche Forschungsgemeinschaft
(Bonn) and Intas (Brussels). The constant interest, deep comments and fruitful 
criticism of Friedrich Hehl are gratefully acknowledged. 

\section{Appendix 1: Irreducible decompositions}\label{irrdecomp}

At first, we recall that the torsion 2-form can be decomposed into the three 
irreducible pieces, $T^{\alpha}={}^{(1)}T^{\alpha} + {}^{(2)}T^{\alpha} + 
{}^{(3)}T^{\alpha}$, where
\begin{eqnarray}
{}^{(2)}T^{\alpha}&=& {1\over 3}\vartheta^{\alpha}\wedge (e_\nu\rfloor 
T^\nu),\label{iT2}\\
{}^{(3)}T^{\alpha}&=& -\,{1\over 3}{}^*(\vartheta^{\alpha}\wedge{}^*
(T^{\nu}\wedge\vartheta_{\nu}))= {1\over 3}e^\alpha\rfloor(T^{\nu}\wedge
\vartheta_{\nu}),\label{iT3}\\
{}^{(1)}T^{\alpha}&=& T^{\alpha}-{}^{(2)}T^{\alpha} - {}^{(3)}T^{\alpha}.
\label{iT1}
\end{eqnarray}

The Riemann-Cartan curvature 2-form is decomposed $R^{\alpha\beta} = 
\sum_{I=1}^6\,{}^{(I)}R^{\alpha\beta}$ into the 6 irreducible parts 
\begin{eqnarray}
{}^{(2)}R^{\alpha\beta} &=& -\,{}^*(\vartheta^{[\alpha}\wedge\Psi^{\beta]}),
\label{curv2}\\
{}^{(3)}R^{\alpha\beta} &=& -\,{\frac 1{12}}\,{}^*(X\,\vartheta^\alpha\wedge
\vartheta^\beta),\label{curv3}\\
{}^{(4)}R^{\alpha\beta} &=& -\,\vartheta^{[\alpha}\wedge\Phi^{\beta]},
\label{curv4}\\
{}^{(5)}R^{\alpha\beta} &=& -\,{\frac 12}\vartheta^{[\alpha}\wedge e^{\beta]}
\rfloor(\vartheta^\alpha\wedge W_\alpha),\label{curv5}\\
{}^{(6)}R^{\alpha\beta} &=& -\,{\frac 1{12}}\,W\,\vartheta^\alpha\wedge
\vartheta^\beta,\label{curv6}\\
{}^{(1)}R^{\alpha\beta} &=& R^{\alpha\beta} -  
\sum\limits_{I=2}^6\,{}^{(I)}R^{\alpha\beta},\label{curv1}
\end{eqnarray}
where 
\begin{equation}
W^\alpha := e_\beta\rfloor R^{\alpha\beta},\quad W := e_\alpha\rfloor W^\alpha,
\quad X^\alpha := {}^*(R^{\beta\alpha}\wedge\vartheta_\beta),\quad X := 
e_\alpha\rfloor X^\alpha,\label{WX}
\end{equation}
and 
\begin{eqnarray}
\Psi_\alpha &:=& X_\alpha - {\frac 14}\,\vartheta_\alpha\,X - {\frac 12}
\,e_\alpha\rfloor (\vartheta^\beta\wedge X_\beta),\label{Psia}\\
\Phi_\alpha &:=& W_\alpha - {\frac 14}\,\vartheta_\alpha\,W - {\frac 12}
\,e_\alpha\rfloor (\vartheta^\beta\wedge W_\beta)\label{Phia}.
\end{eqnarray}

The curvature tensor $R_{\mu\nu\alpha}{}^\beta$ is constructed from the 
components of the 2-form $R_\alpha{}^\beta = {\frac 12}R_{\mu\nu\alpha}
{}^\beta\,\vartheta^\mu\wedge\vartheta^\nu$. The Ricci tensor is defined
as ${\rm Ric}_{\alpha\beta} := R_{\gamma\alpha\beta}{}^\gamma$. The 
curvature scalar $R = g^{\alpha\beta}{\rm Ric}_{\alpha\beta}$ determines
the 6-th irreducible part (\ref{curv6}) since $W \equiv R$. The first 
irreducible part (\ref{curv1}) introduces the generalized Weyl tensor
$C_{\mu\nu\alpha}{}^\beta$ via the expansion of the 2-form ${}^{(1)}
R_\alpha{}^\beta = {\frac 12}C_{\mu\nu\alpha}{}^\beta\,\vartheta^\mu
\wedge\vartheta^\nu$. From (\ref{Phia}) we learn that the 4-th part of
the curvature is given by the symmetric traceless Ricci tensor,   
\begin{equation}
\Phi_\alpha = \left(R_{(\alpha\beta)} - {\frac 14}\,R\,g_{\alpha\beta}
\right)\vartheta^\beta.\label{Ric}
\end{equation}
Accordingly, the 1-st, 4-th and 6-th curvature parts generalize the 
well-known irreducible decomposition of the Riemannian curvature tensor.
The 2-nd, 3-rd and 5-th curvature parts are purely non-Riemannian since
they all arise from the nontrivial right-hand side of the first Bianchi
identity $R_\alpha{}^\beta\wedge\vartheta^\alpha = DT^\beta$, see 
(\ref{WX}) and (\ref{Psia}).

\section{Appendix 2: Proof of Lemmas}\label{prooflemmas}

{\bf Proof of Lemma A}: Consider a chain of identical transformations 
for the 4-form
\begin{eqnarray}
\left({\frac 1 2}\eta^{\alpha\beta\mu\nu}\vartheta_\alpha\wedge A_\beta
\right)\wedge\vartheta_\rho\wedge\vartheta_\sigma &=&{\frac 1 2}
\eta^{\alpha\beta\mu\nu}\eta_{\alpha\rho\sigma\lambda}\eta^\lambda
\wedge A_\beta \nonumber\\
&=&{\frac 1 2}(-\delta^\mu_\sigma\eta^\nu\wedge A_\rho +
\delta^\nu_\sigma\eta^\mu\wedge A_\rho \nonumber\\
&&\quad + \delta^\mu_\rho\eta^\nu\wedge A_\sigma -
\delta^\nu_\rho\eta^\mu\wedge A_\sigma)
-\delta^{[\mu}_\rho\delta^{\nu]}_\sigma\,\eta^\beta\wedge A_\beta \nonumber\\
&=&{\frac 1 2}(\delta^\mu_\sigma\eta_\rho\wedge A^\nu -
\delta^\nu_\sigma\eta_\rho\wedge A^\mu -\delta^\mu_\rho\eta_\sigma\wedge 
A^\nu + \delta^\nu_\rho\eta_\sigma\wedge A^\mu)\nonumber\\
&=&\vartheta^{[\mu}\wedge\eta_{\rho\sigma}\wedge A^{\nu]} 
=\vartheta^{[\mu}\wedge A^{\nu]}\wedge{}^\star\!\left(\vartheta_\rho
\wedge\vartheta_\sigma\right)\nonumber\\
&=&\,{}^\star\!\left(\vartheta^{[\mu}\wedge A^{\nu]}
\right)\wedge\vartheta_\rho\wedge\vartheta_\sigma .
\end{eqnarray}
Comparing the beginning and the end, by Cartan's lemma, one finds the
identity (\ref{idA}). In this calculation we used (\ref{condA}), the
identity for transvection of two Levi-Civita tensors, and the identity
$\vartheta^{\mu}\wedge\eta_{\rho\sigma}=\delta^\mu_\sigma\eta_\rho -
\delta^\nu_\rho\eta_\sigma$. 
Notice that (\ref{condA}) implies $A_\alpha\wedge\eta^\alpha=0$.

{\bf Proof of Lemma B}: Completely analogously to Lemma A, we have
\begin{eqnarray}
\left({\frac 1 2}\eta^{\alpha\beta\mu\nu}\vartheta_\alpha\wedge B_\beta
\right)\wedge\vartheta_\rho\wedge\vartheta_\sigma &=& {\frac 1 2}
\eta^{\alpha\beta\mu\nu}\eta_{\alpha\rho\sigma\lambda}\eta^\lambda
\wedge B_\beta \nonumber\\
&=&{\frac 1 2}(-\delta^\mu_\sigma\eta^\nu\wedge B_\rho + 
\delta^\nu_\sigma\eta^\mu\wedge B_\rho \nonumber\\
&&\quad + \delta^\mu_\rho\eta^\nu\wedge B_\sigma -
\delta^\nu_\rho\eta^\mu\wedge B_\sigma)
-\delta^{[\mu}_\rho\delta^{\nu]}_\sigma\,\eta^\beta\wedge B_\beta \nonumber\\
&=&{\frac 1 2}(-\delta^\mu_\sigma\eta_\rho\wedge B^\nu +
\delta^\nu_\sigma\eta_\rho\wedge B^\mu + \delta^\mu_\rho\eta_\sigma\wedge 
B^\nu - \delta^\nu_\rho\eta_\sigma\wedge B^\mu)\nonumber\\
&=&-\,\vartheta^{[\mu}\wedge\eta_{\rho\sigma}\wedge B^{\nu]}
= -\,\vartheta^{[\mu}\wedge B^{\nu]}\wedge{}^\star\!\left(\vartheta_\rho
\wedge\vartheta_\sigma\right)\nonumber\\
&=&\, -\,{}^\star\!\left(\vartheta^{[\mu}\wedge B^{\nu]}
\right)\wedge\vartheta_\rho\wedge\vartheta_\sigma .
\end{eqnarray}
We thus find the identity (\ref{idB}).

\section{Appendix 3: Proof of geometrical identities}\label{proofgeo}

Here we prove two important geometrical identities for contortion and torsion:
\begin{eqnarray}
{\frac 1 2}\,K^{\mu\nu}\wedge\eta_{\alpha\mu\nu}&\equiv& {}^*(-{}^{(1)}
T_\alpha + 2{}^{(2)}T_\alpha + {\frac 1 2}{}^{(3)}T_\alpha),\label{ID1}\\
D\eta_{\alpha\beta}&\equiv& \vartheta_{[\alpha}\wedge K^{\mu\nu}\wedge
\eta_{\beta]\mu\nu}.\label{ID2}
\end{eqnarray}

Hereafter we use the identity 
\begin{equation}
{}^*(\Phi\wedge\vartheta^\alpha)\equiv e^\alpha\rfloor{}^*\Phi,
\end{equation}
valid for any form $\Phi$. Next, we need the product 
\begin{equation}
\vartheta^\beta\wedge\eta_{\alpha\mu\nu}= \delta^\beta_\alpha\,
\eta_{\mu\nu} + \delta^\beta_\mu\,\eta_{\nu\alpha} + \delta^\beta_\nu\,
\eta_{\alpha\mu}.\label{idve}
\end{equation}

Let us now compute $K^{\mu\nu}\wedge\eta_{\alpha\mu\nu}$. Using 
(\ref{contor}), we find:
\begin{equation}
K^{\mu\nu}\wedge\eta_{\alpha\mu\nu}=-(e^\mu\rfloor T^\nu)\wedge
\eta_{\alpha\mu\nu} + {1\over 2}(e^\mu\rfloor e^\nu\rfloor T_\beta)
\vartheta^\beta\wedge\eta_{\alpha\mu\nu}.\label{Ket1}
\end{equation}
In order to calculate the first term, we start with
\begin{eqnarray}
(e^\mu\rfloor T^\nu)\wedge\eta_{\mu\nu}&=& e^\mu\rfloor (T^\nu\wedge
\eta_{\mu\nu})=e^\mu\rfloor (\eta_\mu\wedge T)= 
-\eta_\mu\,e^\mu\rfloor T \nonumber\\
&=& -{}^*(\vartheta_\mu\,e^\mu\rfloor T)= -{}^*T,
\end{eqnarray}
where $T:=e_\nu\rfloor T^\nu$, and we used the identity $0\equiv e_\nu\rfloor 
(T^\nu\wedge\eta_\mu)=T\wedge\eta_\mu + T^\nu\wedge\eta_{\mu\nu}$. Applying
the interior product $e_\alpha\rfloor$, we find
\begin{equation}
(e_\alpha\rfloor e^\mu\rfloor T^\nu)\,\eta_{\mu\nu} -(e^\mu\rfloor T^\nu)
\wedge\eta_{\alpha\mu\nu} = - e_\alpha\rfloor{}^*T,
\end{equation}
and thus the first term on the right hand side of (\ref{Ket1}) reads
\begin{eqnarray}
-(e^\mu\rfloor T^\nu)\wedge\eta_{\alpha\mu\nu}&\equiv& - e_\alpha\rfloor
{}^*T -(e_\alpha\rfloor e^\mu\rfloor T^\nu)\,\eta_{\mu\nu} \nonumber\\
&=& -{}^*(T\wedge\vartheta_\alpha) - {}^*(\vartheta_\mu\wedge\vartheta_\nu\,
e_\alpha\rfloor e^\mu\rfloor T^\nu) \nonumber\\
&=& {}^*(\vartheta_\alpha\wedge T - T_\alpha
+ e_\alpha\rfloor (\vartheta^\nu\wedge T_\nu)).\label{Ket2}
\end{eqnarray}
The second term on the right hand side of (\ref{Ket1}) is easily computed
with the help of (\ref{idve}): 
\begin{eqnarray}
{1\over 2}(e^\mu\rfloor e^\nu\rfloor T_\beta)
\vartheta^\beta\wedge\eta_{\alpha\mu\nu}&\equiv& {1\over 2}{}^*\!(
\vartheta_\mu\wedge\vartheta_\nu\,e^\mu\rfloor e^\nu\rfloor T_\alpha
\nonumber\\ && -\vartheta_\nu\wedge\vartheta_\alpha\,e^\nu\rfloor T + 
\vartheta_\alpha\wedge\vartheta_\mu\,e^\mu\rfloor T)\nonumber\\
&=& {}^*(-T_\alpha + \vartheta_\alpha\wedge T).\label{Ket3}
\end{eqnarray}
Collecting (\ref{Ket2}) and (\ref{Ket3}) together, we find:
\begin{equation}
K^{\mu\nu}\wedge\eta_{\alpha\mu\nu}\equiv {}^*(-2T_\alpha +2\vartheta_\alpha
\wedge T + e_\alpha\rfloor (\vartheta^\nu\wedge T_\nu)).
\end{equation}
Substituting the definitions (\ref{iT2})-(\ref{iT1}), one proves the
identity (\ref{ID1}).

The proof of the second identity (\ref{ID2}) is more simple. Using the
decomposition (\ref{gagaK}) and (\ref{zeroeta}), we find for the left hand
side:
\begin{equation}
D\eta_{\alpha\beta}=  - K_\alpha{}^\gamma\wedge\eta_{\gamma\beta} -
K_\beta{}^\gamma\wedge\eta_{\alpha\gamma}.
\end{equation}
However, from (\ref{idve}) we derive
\begin{equation}
\vartheta_{[\alpha}\wedge\eta_{\beta]\mu\nu}=g_{\alpha[\mu}\,\eta_{\nu]\beta}
- g_{\beta[\mu}\,\eta_{\nu]\alpha},
\end{equation}
and hence for the right hand side one finds
\begin{equation}
\vartheta_{[\alpha}\wedge K^{\mu\nu}\wedge\eta_{\beta]\mu\nu}=
-K^{\mu\nu}\wedge\vartheta_{[\alpha}\wedge\eta_{\beta]\mu\nu}=
-2\,K_{[\alpha}{}^{\nu}\wedge\eta_{\nu|\beta]},
\end{equation}
which proves (\ref{ID2}).


\begin{thebibliography}{99}

\bibitem{Baekler1}
P. Baekler, {\it Prolongation structure and Backlund transformations of
gravitational double duality equations}, {\sl Class. Quantum Grav.} {\bf 8}
(1991) 1023-1046.

\bibitem{Baekler2}
P. Baekler and F.W. Hehl, {\it On the dynamics of the torsion of space-time: 
exact solutions in a gauge theoretical model of gravity}, in: {\sl ``From 
$SU(3)$ to Gravity. Festschrift in honor of Y.Ne'eman" /Eds. E. Gotsman,
and G. Tauber} (Cambridge Univ. Press: Cambridge, 1985) 341-359.

\bibitem{Baekler3}
P. Baekler, F.W. Hehl, and H.J. Lenzen, {\it Vacuum solutions with double
duality properties of the Poincar\'e gauge field theory II}, in:
{\sl ``Proc. 3rd Marcel Grossmann Meeting on General Relativity"
/Ed. Hu Ning} (North Holland: Amsterdam, 1983) 107-128.

\bibitem{Baekler4}
P. Baekler, F.W. Hehl, and E.W. Mielke, {\it Vacuum solutions with
double duality properties of a quadratic Poincar\'e gauge field
theory}, in: {\sl ``Proc. of the 2nd Marcel Grossmann Meeting on
Recent Progress of the Fundamentals of General Relativity"
/Ed. R. Ruffini} (North Holland: Amsterdam, 1981) 413-450.

\bibitem{Baekler5}
P. Baekler and E.W. Mielke,  {\it Effective  Einsteinian  gravity  from
Poincar\'e gauge field theory}, {\sl Phys. Lett.} {\bf A113} (1986) 471-475.

\bibitem{Baekler6}
P. Baekler and E.W. Mielke, {\it Hamiltonian structure of the Poincar\'e gauge
theory and separation of non-dynamical variables in exact torsion solutions},
{\sl Fortschr. Phys.} {\bf 36} (1988) 549-594.

\bibitem{Baekler7}
P. Baekler, E.W. Mielke, R. Hecht, and F.W. Hehl, {\it Kinky torsion in a
Poincar\'e gauge model of gravity coupled to a massless  scalar
field}, {\sl Nucl. Phys.} {\bf B288} (1987) 800-812.

\bibitem{Benn}
I.M. Benn, T. Dereli, and R.W. Tucker, {\it Double-dual solutions of
generalized theories of gravitation}, {\sl Gen. Relat. Grav.} {\bf 13}
(1981) 581-589.

\bibitem{Blag}
M. Blagojevi\'c, {\it Gravitation and Gauge Symmetries} (IOP Publishing: 
Bristol, 2002).


\bibitem{Debney}
G. Debney, E.E. Fairchild,Jr., and S.T.C. Siklos,  {\it Equivalence of
vacuum Yang-Mills gravitation and vacuum Einstein  gravitation},
{\sl Gen. Relat. Grav.} {\bf 9} (1978) 879-887.

\bibitem{Fair1}
E.E. Fairchild, Jr., {\it Gauge theory of gravitation}, {\sl Phys. Rev.}
{\bf D14} (1976) 384-391; (E) 3439.

\bibitem{Fair2}
E.E. Fairchild, Jr., {\it Yang-Mills formulation of gravitational
dynamics}, {\sl Phys. Rev.} {\bf D16} (1977) 2438-2447.

\bibitem{Frolov}
B.N. Frolov, {\it Tetrad Palatini formalism and quadratic Lagrangians in the
gravitational field theory}, {\sl Acta Phys. Pol.} {\bf B9} (1978) 823-829.

\bibitem{Garecki}
J. Garecki, {\it Gauge theory of gravitation with quadratic Lagrangian 
$L_{g}=\alpha\Omega^{i}_{\ j}\wedge\eta_{i}^{\ j} + 
\beta\Omega_{\ j}^{i}\wedge\ast\Omega_{\ i}^{j} + 
\gamma\Theta^{i}\wedge\ast \Theta_{i}$ containing Einsteinian term
$\alpha\Omega^{i}_{\ j}\wedge\eta_{i}^{\ j}$ and spherically symmetric
solutions to its field equations}, in: {\sl ``On Relativity Theory. Proc. of 
Sir A.Eddington Centenary Symp." /Eds. Y.Choquet-Bruhat and T.M.Karade} 
(World Scientific: Singapore, 1985) vol. {\bf 2}, 232-260.

\bibitem{Glad}
M.S. Gladchenko and V.V. Zhytnikov, {\it Post-Newtonian effects in the
quadratic Poincar\'e gauge theory of gravitation}, {\sl Phys. Rev.}
{\bf D50} (1994) 5060-5071.

\bibitem{Hecht}
R.D. Hecht, J. Lemke, and R.P. Wallner, {\it Can Poincar\'e gauge theory be 
saved?}, {\sl Phys. Rev.} {\bf D44} (1991) 2442-2451.

\bibitem{Hehl1}
F.W. Hehl, P. von der Heyde, G.D. Kerlick, and J.M. Nester, {\it General 
relativity with spin and torsion: foundation and prospects}, {\sl Revs. 
Mod. Phys.} {\bf 48} (1976) 393-416.

\bibitem{Hehl2}
F.W. Hehl, J.D. McCrea, E.W. Mielke, and Y. Ne'eman, {\it Metric-affine gauge
theory of gravity: field equations, Noether identities, world spinors, and
breaking of dilaton invariance}, {\sl Phys. Repts.} {\bf 258} (1995) 1-171.

\bibitem{Hehl3}
F.W. Hehl, J. Nitsch, and P. von der Heyde, {\it Gravitation and the Poincar\'e 
gauge field theory with quadratic Lagrangians}, in: {\sl ``General Relativity 
and Gravitation: One Hundred Years after the Birth of Albert Einstein" 
/Ed. A.Held} (Plenum:  New  York, 1980) vol. {\bf 1}, 329-355.

\bibitem{Birkbook}
F.W.~Hehl and Yu.N.~Obukhov, {\it Foundations of Classical Electrodynamics 
--- Charge, Flux, and Metric}. (Birkh{\"a}user: Boston, 2003).

\bibitem{Hehl4}
F.W. Hehl, A. Mac\'{\i}as, E.W. Mielke, and Yu.N. Obukhov, {\it On the 
structure of the energy-momentum and the spin currents in Dirac's electron 
theory}, in: {\sl ``On Einstein's path'', Essays in honor of E.Schucking},
Ed. A. Harvey (Springer: New York, 1998) 257-274.


\bibitem{Ivan1}
D.D. Ivanenko, P.I. Pronin, and G.A. Sardanashvily, {\it Gauge theory of
gravity} (Moscow University Publ. House: Moscow, 1985) 144  p. (in Russian).

\bibitem{Ivan2}
D. Ivanenko and G. Sardanashvily, {\it The gauge treatment of gravity},
{\sl Phys. Repts.} {\bf 94} (1983) 3-45.

\bibitem{Katanaev}
M.O. Katanaev, {\it Kinematic part in the  dynamical  torsion  theory},
{\it Theor. Math. Phys.} {\bf 72} (1987) 735-741 [{\sl Theor. Math. Phys.} 
{\bf 72} (1987) 79-88 (in Russian)].

\bibitem{Kopc}
W. Kopczy\'nski,  {\it Problems  with  metric-teleparallel  theories  of
gravitation}, {\sl J. Phys.} {bf A15} (1982) 493-506.

\bibitem{Kuhfuss}
R. Kuhfuss and J. Nitsch, {\it Propagating modes in gauge field theories
of gravity}, {\sl Gen. Relat. Grav.} {\bf 18} (1986) 1207-1227.

\bibitem{Leclerc1}
M. Leclerc, {\it Teleparallel limit of Poincar\'e gauge theory}, 
{\sl Phys. Rev.} {\bf D71} (2005) 027503.

\bibitem{Leclerc2}
M. Leclerc, {\it One-parameter teleparallel limit of Poincar\'e gravity}, 
{\sl Phys. Rev.} {D72} (2005) 044002. 

\bibitem{Mccrea1}
J.D. McCrea, {\it The use of REDUCE in finding exact solutions  of  the
quadratic Poincar\'e  gauge  field  equations},  in:  {\sl ``Classical
General  Relativity"  /  Eds.  W.B.Bonnor,   J.N.Islam,   and
M.A.H.MacCallum} (Cambridge Univ. Press: Cambridge, 1984) 173-182.

\bibitem{Mccrea2}
J.D. McCrea, {\it Poincar\'e gauge theory of gravitation: foundations, exact
solutions and computer algebra}, in: {\sl ``Proc. of the 14th Intern. Conf. 
on Differential Geometric Methods in Mathematical Physics, Salamanca, 1985}
/Eds. P.L. Garcia and A. P\'erez-Rend\'on, {\sl Lect. Notes Math.} {\bf 1251}
(1987) 222-237.

\bibitem{Mccrea3}
J.D. McCrea, {\it REDUCE in general relativity and Poincar\`e gauge theory},
in: {\sl ``Algebraic computing in general relativity'', Lect. Notes of the
1st Brasil. School on Comp. Algebra, Rio de Janeiro, July-August 1989}
/Eds. M.J. Rebou\c{c}as and W.L. Roque (Clarendon Press: Oxford, 1994)  
173-263.

\bibitem{Mielke1}
E.W. Mielke, {\it Reduction of the Poincar\'e gauge  field  equations  by
means of duality rotations}, {\sl J. Math. Phys.} {\bf 25} (1984) 663-668.

\bibitem{Mielke2}
E.W. Mielke, {\it On pseudoparticle solutions  in  the  Poincar\'e  gauge
theory of gravity}, {\sl Fortschr. Phys.} {\bf 32} (1984) 639-660.

\bibitem{Mielke3}
E.W. Mielke, {\it Geometrodynamics of gauge fields - On the geometry of 
Yang-Mills and gravitational gauge theories} (Akademie Verlag: Berlin, 1987).

\bibitem{Mielke4}
E.W. Mielke and R.P. Wallner, {\it Mass and spin of double dual solutions in
Poincar\'e gauge theory}, {\sl Nuovo Cim.} {\bf B101} (1988) 607-624.

\bibitem{Mink}
A.V. Minkevich, {\it Problem of cosmological singularity and gauge
theories of gravitation}, {\sl Acta Phys. Pol.} {\bf 29} (1998) 949-960.

\bibitem{Nester}
J.M. Nester, {\it Is there really a problem with the teleparallel theory?},
{\sl Class. Quantum Grav.} {\bf 5} (1988) 1003-1010.

\bibitem{Neville}
D.E. Neville, {\it Gravity theories with propagating torsion}, 
{\sl Phys. Rev.} {\bf D21} (1980) 867-873.

\bibitem{Fermion}
Yu.N. Obukhov, {\it On gravitational interaction of fermions},
{\sl Fortschritte der Physik} {\bf 50} (2002) 711-716.

\bibitem{OPZ}
Yu.N. Obukhov, V.N. Ponomariev, and V.V. Zhytnikov, {\it Quadratic Poincar\'e
gauge theory of gravity: a comparison with the general relativity theory},
{\sl Gen. Relat. Grav.} {\bf 21} (1989) 1107-1142.

\bibitem{Percacci}
R. Percacci, {\it Geometry of nonlinear field theories} (World Scientific:
Singapore, 1986) 255 p.

\bibitem{PBO}
V.N. Ponomariev, A.O. Barvinski, and Yu.N. Obukhov, {\it Geometrodynamical
methods and the gauge approach to the theory of  gravitational interactions}
(Energoatomizdat: Moscow,  1985)  168  p. (in Russian).

\bibitem{Rama}
S. Ramaswamy and P. Yasskin, {\it Birkhoff theorem for an $R+R^2$ theory
of gravity with torsion}, {\sl Phys. Rev.} {\bf D19} (1979) 2264-2267.

\bibitem{Rauch}
R.T. Rauch, S.J. Shaw, and H.T. Nieh, {\it Birkhoff's theorem for ghost-free
tachyon-free $R+R^{2}+Q^2$ theories with torsion}, {\sl Gen.
Relat. Grav.}  {\bf 14} (1982) 331-354.

\bibitem{Sard}
G. Sardanashvily and O. Zakharov, {\it Gauge Gravitation Theory},
(World Scientific: Singapore, 1992).


\bibitem{Sezgin1}
E. Sezgin and P. van Nieuwenhuizen, {\it New ghost-free gravity  Lagrangians 
with propagating torsion}, {\sl Phys. Rev.} {\bf D21} (1980) 3269-3280.

\bibitem{Sezgin2}
E. Sezgin, {\it A class of ghost-free gravity Lagrangians with  massive or
massless propagating torsion}, {\sl Phys. Rev.} {\bf D24} (1981) 1677-1680.

\bibitem{Shapiro}
I.L. Shapiro, {\it Physical aspects of the space-time torsion}, 
{\sl Phys. Repts.} {\bf 357} (2002) 113-213. 

\bibitem{Szczyrba}
W. Szczyrba, {\it Dynamics of quadratic Lagrangians in gravity. 
Fairchild's theory}, {\sl Contemp. Math.} {\bf 71} (1988) 167-180.

\bibitem{Trautman1}
A. Trautman, {\it Recent advances  in  the  Einstein-Cartan  theory  of
gravity}, {\sl Ann. N.Y. Acad. Sci.} {\bf 262} (1975) 241-245.

\bibitem{Trautman2}
A. Trautman, {\it Fiber bundles, gauge fields and gravitation}, in: {\sl 
``General Relativity and Gravitation: One Hundred Years after the Birth of
 Albert Einstein" /Ed. A.~Held} (Plenum: New York, 1980) vol. {\bf 1}, 287-308.

\bibitem{Trautman3}
A. Trautman, {\it Differential geometry  for  physicists}  (Bibliopolis:
Naples, 1984).

\bibitem{Tres}
R. Tresguerres and E.W. Mielke, {\it Gravitational Goldstone fields from 
affine gauge theory}, {\sl Phys. Rev.} {\bf D62} (2000) 044004 (7 pages). 

\bibitem{Tseytlin}
A.A. Tseytlin, {\it Poincar\'e and de Sitter gauge  theories  of  gravity
with propagating torsion}, {\sl Phys. Rev.} {\bf D26} (1982) 3327-3341.

\bibitem{Wallner}
R.P. Wallner, {\it Exact solutions in $U_4$ gravity. I. The ansatz for self 
double duality curvature}, {\sl Gen. Relat. Grav.} {\bf 23} (1991) 623-639.

\bibitem{Vadim1}
V.V. Zhytnikov, {\it Wavelike exact solutions of $R+R^{2}+Q^{2}$ gravity},
{\sl J. Math. Phys.} {\bf 35} (1994) 6001-6017.

\bibitem{Vadim2}
V.V. Zhytnikov, {\it Double duality and hidden gauge freedom in the
Poincar\'e gauge theory of gravitation}, {\sl Gen. Rel. Grav.} {\bf 28}
(1996) 137-162.

\bibitem{Vadim3}
V.V. Zhytnikov and V.N. Ponomariev, {\it On the equivalence of vacuum 
equations in quadratic gauge theory of gravity and in general relativity}, in:
{\sl ``Probl. Theor. Grav. and Elem. Particles" /Ed. K.P. Stanyukovich}
(Energoatomizdat: Moscow, 1986) {\bf 17}, 93-101 (in Russian).



\end{thebibliography}
\end{document}